\documentclass[pra,aps,twocolumn]{revtex4}
\usepackage{graphicx}
\usepackage{amsmath}
\usepackage{bm}

\newcommand{\be}{\begin{equation}}
\newcommand{\ee}{\end{equation}}
\newcommand{\bea}{\begin{eqnarray}}
\newcommand{\eea}{\end{eqnarray}}
\newcommand{\bei}{\begin{itemize}}
\newcommand{\eei}{\end{itemize}}

\begin{document}

\title{Topological $p_x+ip_y$ Superfluid Phase of Fermionic Polar Molecules}

\author{J. Levinsen$^{1,2}$, N. R. Cooper$^{1,2}$, and G. V. Shlyapnikov$^{2,3,4}$}
\affiliation{\mbox{$^{1}$T.C.M. Group, University of Cambridge, Cavendish Laboratory, J.J. Thomson Ave., Cambridge CB3 0HE, UK}\\
\mbox{$^{2}$Laboratoire de
  Physique Th\'eorique et Mod\`eles Statistiques, CNRS and Universit\'e
  Paris Sud, UMR8626, 91405 Orsay, France}\\
\mbox{$^{3}$Van der Waals-Zeeman Institute, University of Amsterdam, Science Park 904, 1098 XH Amsterdam, The Netherlands}\\
\mbox{$^4$ Kavli Institute for Theoretical Physics, University of California, Santa Barbara, California 93106-4030, USA}}
\begin{abstract}

We discuss the topological $p_x+ip_y$ superfluid phase in a 2D gas of single-component fermionic polar molecules dressed by a circularly polarized microwave
field. This phase emerges because the molecules may interact with each other via a potential $V_0(r)$ that has an attractive dipole-dipole $1/r^3$ tail, which provides
$p$-wave superfluid pairing at fairly high temperatures. We calculate the amplitude of elastic $p$-wave scattering in the potential $V_0(r)$ taking into
account both the anomalous scattering due to the dipole-dipole tail and the short-range contribution. This amplitude is then used for the analytical and numerical solution of the
renormalized BCS gap equation which includes the second order Gor'kov-Melik-Barkhudarov corrections and the correction related to the effective mass of the quasiparticles. We 
find that the critical temperature $T_c$ can be varied within a few orders of magnitude by modifying the short-range part of the potential $V_0(r)$. The decay of the system via
collisional relaxation of molecules to dressed states with lower energies is rather slow due to the necessity of a large momentum transfer. The presence of a constant transverse
electric field reduces the inelastic rate, and the lifetime of the
system can be of the order of seconds even at 2D densities $\sim 10^9$
cm$^{-2}$. This leads to $T_c$ of up to a
few tens of nanokelvins and makes it realistic to obtain the topological $p_x+ip_y$ phase in experiments with ultracold polar molecules. 

\end{abstract}
\date{\today}
\maketitle

\section{Introduction}

The recent breakthrough in creating ultracold diatomic polar molecules
in the ground ro-vibrational state \cite{Ni2008,Deiglmayr2008} and
cooling them towards quantum degeneracy \cite{Ni2008} has opened
fascinating prospects for the observation of novel quantum phases
\cite{Baranov2008,Lahaye2009,Pupillo2008,Wang2006,Buchler2007,Cooper2009,Pikovski2010}.
A serious problem in this direction is related to ultracold chemical
reactions, such as KRb+KRb$\Rightarrow$K$_2$+Rb$_2$ observed in the JILA
experiments with KRb molecules \cite{Ospelkaus2010}, which places
severe limitations on the achievable density in three-dimensional
samples. In order to suppress chemical reactions and perform
evaporative cooling it has been proposed to induce a strong
dipole-dipole repulsion between the molecules by confining them to a
(quasi)two-dimensional (2D) geometry and orienting their dipole
moments (by a strong electric field) perpendicularly to the plane of
the 2D translational motion \cite{Ni2010}.  Nevertheless, in order to
prevent the approach of colliding molecules at short separations,
where the chemical reactions occur, and to proceed with evaporative
cooling, one has to have a fairly strong confinement to the 2D regime
\cite{Bohn1,Baranov2010,Bohn2}.  The suppression of chemical reactions
by nearly two orders of magnitude in the quasi2D geometry has been demonstrated
in the recent JILA experiment \cite{Miranda2011}. It should be noted, however, that these reactions are not expected to
occur for all polar molecules of alkali atoms \cite{Hutson}, on which
experimental efforts are presently focused.  While present for KRb
molecules and for the molecules containing a Li atom, they are
energetically unfavorable, for example, in the case of NaK
molecules. We thus expect that future investigations of many-body
physics will deal either with molecules which do not undergo
ultracold chemical reactions, or otherwise are strongly confined to
the 2D regime.

One of the challenging goals in the studies of many-body physics is
the creation of the topological $p_x+ip_y$ superfluid phase for
identical fermions in two dimensions (2D) (see \cite{Radzihovsky} for
a review). This phase, first discussed in relation to superfluid $^3$He
and the fractional quantum Hall effect \cite{Volovik,Green}, has
exotic topological properties at positive chemical potential $\mu>0$,
i.e. in the BCS regime. The topological nature of this superfluid
phase supports a gapless Majorana mode
at the boundary to a vacuum, and the quantized vortices in the
superfluid carry local zero energy Majorana modes on their
cores\cite{Green,sternannals}. These Majorana modes are predicted to
cause the vortices to obey non-abelian exchange statistics, which has
potential applications in topologically protected quantum information
processing \cite{nayakrmp}.  There is significant interest in finding
physical realizations of this topological superfluid phase in which
the exotic physics of these Majorana modes can be detected.

The $p_x+ip_y$ topological phase can be the ground state of ultracold
identical fermionic atoms interacting via a short-range potential
\cite{Radzihovsky}. However, away from a $p$-wave Feshbach resonance
the superfluid transition temperature $T_c$ is vanishingly low. On
approach to the resonance it increases but the system becomes unstable
due to the formation of long-lived diatomic quasibound states and
their collisional relaxation into deep molecular states
\cite{Gurarie,Castin}. A stable $p_x+ip_y$ state has been recently
proposed for fermionic polar molecules with a large dipole moment
\cite{Cooper2009}. When these molecules are confined to a 2D geometry
and dressed by a microwave field (MW) which is nearly resonant with
the transition between the lowest and the first excited rotational
levels, they acquire an attractive $1/r^3$ dipole-dipole
interaction. This leads to superfluid pairing of $p_x+ip_y$ symmetry,
and due to the anomalous contribution of the $1/r^3$ tail to the
scattering amplitude, already in the BCS limit the superfluid
transition temperature can be made a sizeable fraction of the Fermi
energy. At the same time, collisional decay processes remain
sufficiently slow to allow the experimental realization of this phase
\cite{Cooper2009}. Several other ways in which a stable $p$-wave
coupled superfluid phase may be obtained have been proposed in recent
years \cite{
  Sarma2008,Cherng2008,Sato2009,Nishida2009,Bulgac2009,Massignan2010,Nishida2010}.

In this paper we present a complete analysis of the BCS limit of the
$p_x+ip_y$ phase of MW-dressed fermionic polar molecules. In
particular, using the many-body perturbative approach up to second
order, we derive a relation for the transition temperature $T_c$. It is
then shown how $T_c$ and the collisional stability may be manipulated
by tuning the short-range part of a MW-induced effective
molecule-molecule interaction potential or by applying a constant
electric field perpendicular to the plane of the translational motion.
Our analysis is confirmed by numerical calculations, which
are also extended to the regime of moderately strong interactions.

The paper is organized as follows. In Section II we discuss a single
polar molecule in the presence of a circularly polarized nearly
resonant microwave field and a constant electric field as shown in
Fig.~1. We then show that two microwave-dressed polar molecules
undergoing a 2D translational motion may interact with each other via
a potential $V_0 (r)$ which has a repulsive core, a potential well,
and an attractive $1/r^3$ tail. In Section III we find the amplitude
of elastic $p$-wave scattering of particles in the potential $V_0(r)$,
which takes into account both the anomalous scattering due to the
$1/r^3$ tail and the short-range contribution. Section IV is dedicated
to the analysis of the decay of the gas due to collisional relaxation
of the molecules to lower dressed states. The presence of a constant
electric field is found to reduce the inelastic rate, and the lifetime
of the system can be of the order of seconds even at 2D densities
$\sim 10^9$ cm$^{-2}$. In Section V we present the renormalized BCS
gap equation which takes into account the second order
Gor'kov-Melik-Barkhudarov processes and the effective mass of
the quasiparticles. In Section VI we obtain analytical and numerical
solutions of the gap equation, show that the ground state has
$p_x+ip_y$ symmetry, and reveal the influence of the short-range part
of the scattering potential $V_0(r)$ on the critical temperature. In
Section VII we conclude, emphasizing that it is realistic to obtain
the topological $p_x+ip_y$ phase for microwave-dressed polar molecules
in the 2D geometry, with a critical temperature of up to a few tens of
nanokelvin.

\section{Interaction potential for MW-dressed polar molecules} 

The microwave dressing of polar molecules has been proposed to tune
the molecule-molecule interaction potential \cite{Micheli} and to form
a repulsive shield for suppressing inelastic losses \cite{Gorshkov}.
The rotational part of the Hamiltonian of a single polar molecule in a
circularly polarized MW-field ${\bf E}_{ac}$ and a constant electric
field ${\bf E}_{dc}$ reads:
\begin{equation}           \label{H0}
\hat H_0=B\hat{\bf J}^2-\hat{\bf d}\cdot({\bf E}_{dc}+{\bf E}_{ac}),
\end{equation}
where $\hat{\bf J}$ is the operator of the rotational moment, $B$ is the
rotational constant of the molecule, and $\hat{\bf d}$ is the operator
of the dipole moment. In the following we assume that ${\bf E}_{dc}$
is parallel to the $z$-axis, and the vector ${\bf E}_{ac}$ rotates in
the $\{x,y\}$ plane (see Fig.~1).

\begin{figure}[ttp]
\includegraphics[width=0.98\columnwidth]{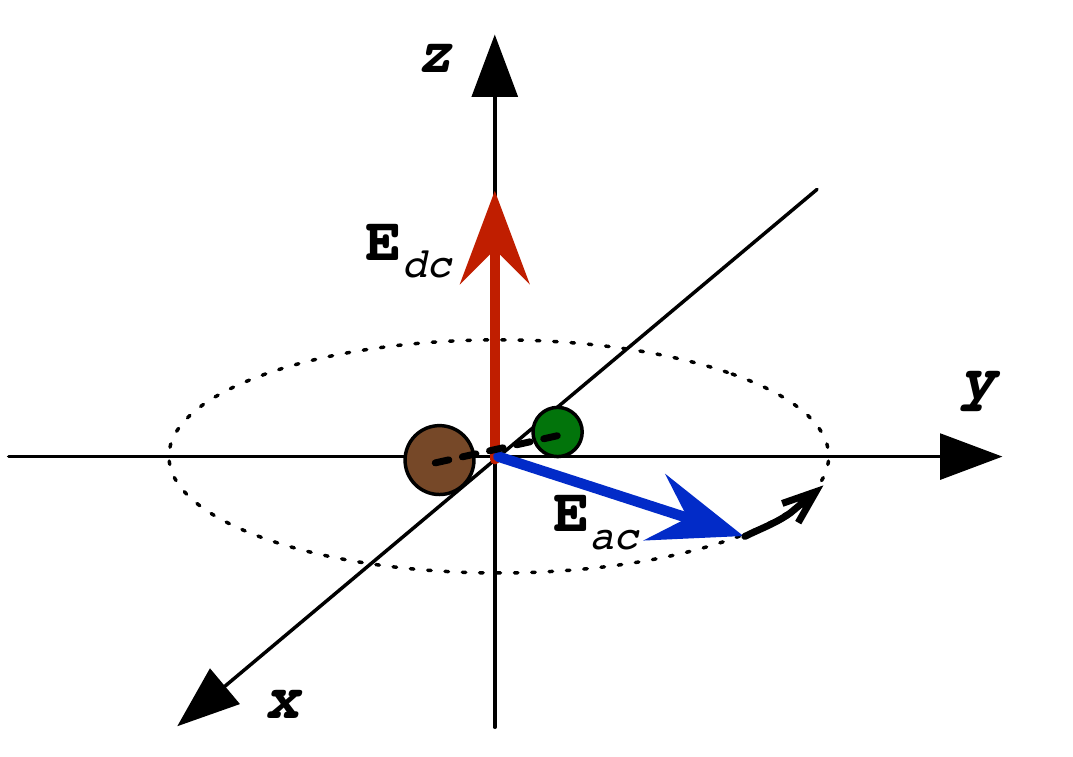}
\caption{(color online). Polar molecule (brown and green circles) in the considered configuration of fields. The {\em ac} MW field rotates
  with frequency $\omega$ in the plane orthogonal to the {\em dc} field 
  \label{fig:fields}}
\end{figure}

In the absence of electric fields, good quantum numbers are the
rotational moment $J$ and its projection on the $z$-axis, $M_J$, and
the diagonal matrix element $\langle J,M_J|\hat{\bf d}|J,M_J\rangle$ is
equal to zero.
The $dc$-field preserves $M_J$ but couples states with different $J$.
We will consider a rather weak $dc$-field
so that the parameter
\begin{equation}          \label{beta}
\beta=\frac{dE_{dc}}{B}
\end{equation}
is small, where $d$ is the molecular dipole moment. Within the
manifold that includes the original states with $J=0$ and $J=1$, the new basic states will be called $|\Phi_{00}\rangle$, 
$|\Phi_{1-1}\rangle$,   $
|\Phi_{10}\rangle$,   and $|\Phi_{11}\rangle$. The transition matrix elements of the operator of the dipole moment, which are needed for further discussions,  are given by
\cite{Micheli}:    
\begin{eqnarray}
&&d_c=|\langle\Phi_{00}|\hat{d}_{\mp}|\Phi_{1\pm 1}\rangle|\approx \frac{d}{\sqrt{3}}\left(1-\frac{49\beta^2}{1440}\right),  \label{dc} \\
&&d_g=|\langle\Phi_{00}|\hat{d}_z|\Phi_{00}\rangle|\approx\frac{d\beta}{3}, \label{dg} \\
&&d_e=|\langle\Phi_{1\pm 1}|\hat{d}_z|\Phi_{1\pm 1}\rangle|\approx \frac{d\beta}{10}, \label{de}
\end{eqnarray} 
where we have omitted terms that are higher order in $\beta$. 

Consider the application of a circularly polarized MW-field with frequency
$\omega$ close to the frequency $\omega_0$ of the transition between
the states $|\Phi_{00}\rangle$ and $| \Phi_{11}\rangle$. For the Rabi
frequency $\Omega_R=d_cE_{ac}/\hbar$ and the frequency detuning
$\delta=\omega-\omega_0$ satisfying the inequality
$|\delta|,\Omega_R\ll\omega_0$, the rotating wave approximation is
valid and the MW-field couples only the states $|\Phi_{00}\rangle$ and
$|\Phi_{11}\rangle$. The resulting eigenstates are:
\begin{eqnarray}
&&|+\rangle=a|\Phi_{00}\rangle+b|\Phi_{11}\rangle, \label{+} \\
&&|-\rangle=b|\Phi_{00}\rangle-a|\Phi_{11}\rangle, \label{-}
\end{eqnarray} 
with $a=-A/\sqrt{A^2+\Omega_R^2}$, $b =
\Omega_R/\sqrt{\Omega_R^2+A^2}$, and $A=(\delta
+\sqrt{\delta^2+4\Omega_R^2})/2$. We will consider $\delta\agt
\Omega_R$ , and choose $\delta > 0$ such that the energy of the state
$|+\rangle$ lies {\it above} the energies of the states $|-\rangle$
and $|\Phi_{1-1}\rangle$.
Ramping the MW field on adiabatically then ensures that the ground
state $|0,0\rangle$ evolves into the state $|+\rangle$. Thus all
molecules can be prepared in this state.

We remark that it was demonstrated in Ref. \cite{Ospelkaus2010b} that
polar molecules may be prepared in any hyperfine substate of the
ground ro-vibrational state. In particular, we shall assume that the
molecules are prepared in the state with maximum projection of the
magnetic moment such that the circularly polarized MW field does not
cause mixing of hyperfine levels.

We now consider two MW-dressed polar molecules undergoing a 2D motion
in the $\{x,y\}$ plane and separated by a distance $r$ which greatly
exceeds the radius of the ordinary van der Waals interaction
potential. The presence of the $dc$ and MW-fields introduces a
dipole-dipole interaction between the molecules:
\begin{equation}         \label{Hd}
\hat H_d=\frac{\hat{\bf d}_1\cdot\hat{\bf d}_2-3(\hat{\bf d}_1\cdot \hat{\bf
    r})(\hat{\bf d}_2\cdot \hat{\bf r})}{r^3}.
\end{equation}
Assuming that the molecules are at a fixed separation ${\bf r}$ we employ the Born-Oppenheimer approximation and determine an effective intermolecular potential
induced by the $dc$ and MW-fields. For this purpose we have to diagonalize the Hamiltonian $\hat H=\hat H_0+\hat H_d$. The dipole-dipole interaction couples the molecular states
$|g\rangle\equiv |\Phi_{00}\rangle$,  $|e\rangle\equiv |\Phi_{11}\rangle$, and $|{\bar e}\rangle\equiv |\Phi_{1-1}\rangle$, whereas the state $|\Phi_{10}\rangle$ remains decoupled
from them. The wavefunction of a two-molecule dressed state can be either symmetric (even parity) or antisymmetric (odd parity) with respect to permutation of the molecules. For
studying
topological superfluids we will be interested in the states of even
parity. We then have a basis of five two-particle states:
\begin{eqnarray} 
|g,g\rangle,\,\frac{(|e,g\rangle\!+\!|g,e\rangle)}{\sqrt{2}},\,\frac{(|{\bar e},g\rangle\!+\!|g,{\bar e}\rangle)}{\sqrt{2}},\,|e,e\rangle,\,\frac{(|{\bar e},e\rangle\!+\!|e,{\bar
e}\rangle)}{\sqrt{2}}, \nonumber
\end{eqnarray}
(the state $|{\bar e},{\bar e}\rangle$ is decoupled from these states). The Hamiltonian acting on the spinor of these five states is represented by a matrix \cite{Gorshkov}:
\begin{widetext}
\begin{equation}            \label{Hmatrix}
\hat H=\hbar\delta\left(\begin{array}{ccccc}
\beta^2/3x^3 & \sqrt2\Omega_R /\delta & 0 & 0 & 0\\
\sqrt2\Omega_R /\delta & \left(\beta^2/10-1/2\right)/x^3-1
& 3/2x^3 & \sqrt2\Omega_R /\delta & 0\\
0 & 3/2x^3 & \left(\beta^2/10-1/2\right)/x^3-1
& 0 & \Omega_R /\delta \\
0 & \sqrt2\Omega_R /\delta & 0 & 3\beta^2/100x^3-2 & 0\\
0 & 0 & \Omega_R /\delta & 0 & 3\beta^2/100x^3-2
\end{array}\right),
\end{equation}
\end{widetext}
where $x\equiv r/r_{\delta}$, $r_{\delta}=(d_c^2/\hbar\delta)^{1/3}$,
and we omit terms that are higher order in $\beta$. Diagonalizing the
matrix (\ref{Hmatrix}) we obtain a number of effective intermolecular
potentials (potential curves) $V_{\mbox{\tiny eff}}(r)$ for MW-dressed
polar molecules, depending on the states of the molecules at an
infinite separation (see Fig.~2).

\begin{figure}[ttp]
\includegraphics[width=1.0\columnwidth]{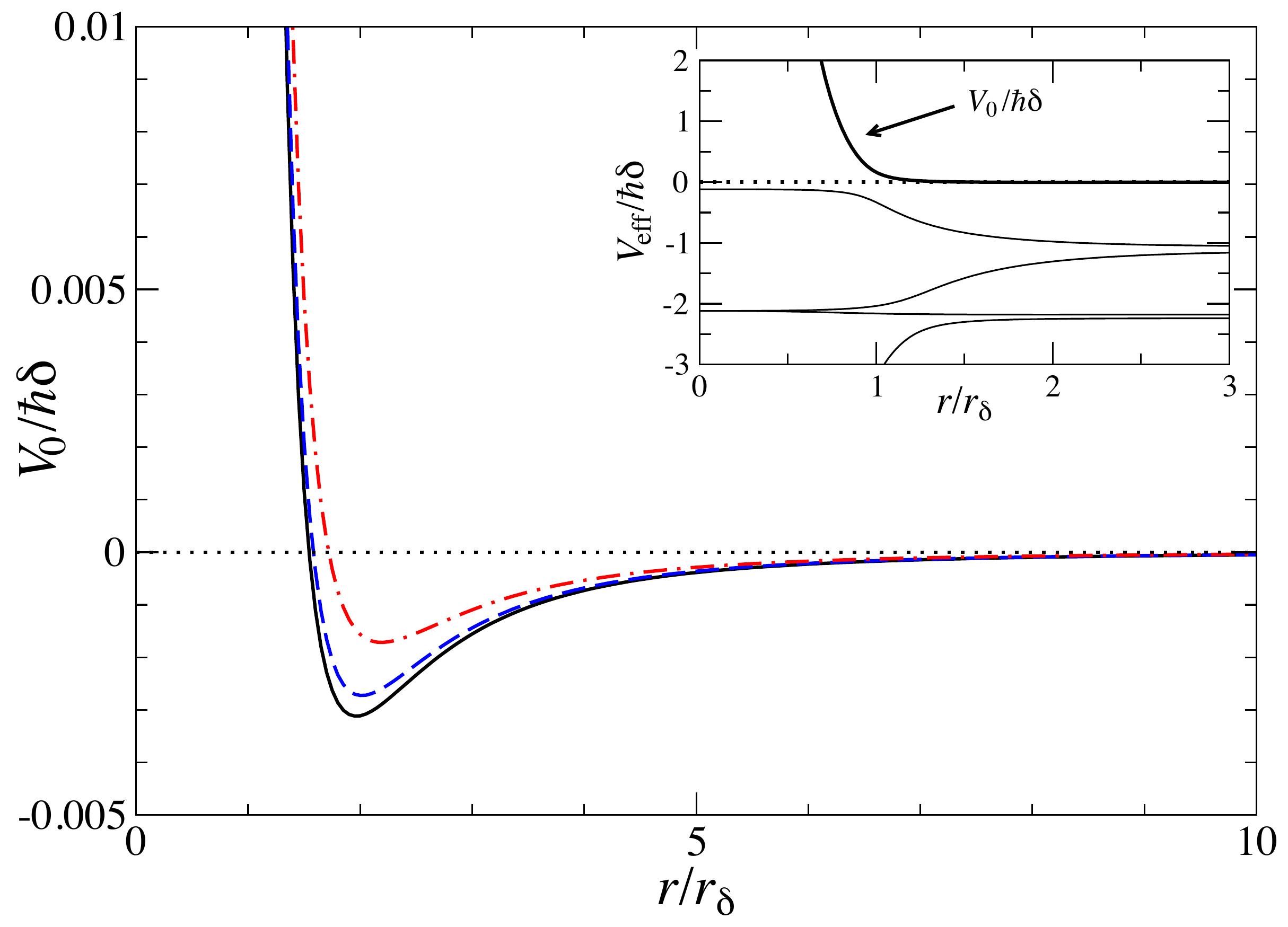}
\caption{ \label{fig:bosurface} (color online). Potential energy curve $V_0(r)$ for
  two $|+\rangle$ state molecules, computed for $\Omega_R = 0.25
  \delta$. The black (solid) curve corresponds to the value of the perpendicular $dc$
  field $\beta=0$, the blue (dashed) curve to $\beta=0.1$, and the red
  (dash-dotted) curve to $\beta=0.2$.  
  Anticrossings of $V_0(r)$  with other field-dressed levels of even parity occur at distances
  $r\sim r_{\delta}$, as shown in the inset for $\Omega_R=0.25\delta$ and $\beta=0$.
} 
\end{figure}

For two molecules which at $r\rightarrow\infty$ are in the state
$|+\rangle$, the long-range tail of such a potential denoted here as
$V_0(r)$ is
\begin{equation}           \label{Vefftail}
V_0(r\to\infty)=-\frac{\hbar^2}{m}\frac{r_*}{r^3},
\end{equation}
with the length scale $r_*$ given by
\begin{eqnarray}            \label{rstar}
&&\hspace{-7mm}
r_*=\frac{md^2}{3\hbar^2}\frac{(\Omega_R /\delta)^2}{1+4(\Omega_R /\delta)^2}
\Big\{1-3\beta^2\times     \nonumber \\
&&\left[49/2160+\left(\frac{7+13\sqrt{1+4(\Omega_R /\delta)^2}}
{60\Omega_R /\delta}\right)^2\right]\Big\}
\end{eqnarray}
and decreasing with an increase in $\beta$. For small
$\Omega_R/\delta$ and sufficiently large $\beta$ the parameter $r_*$
becomes negative, which leads to a repulsive tail of $V_0(r)$. In this
case the $dc$-field dominates over the effect of the MW-dressing, so
that on average the molecules are perpendicular to the $\{x,y\}$ plane
and exhibit a repulsive dipole-dipole interaction at large $r$. In the
following we consider only the case where $r_*$ is positive and the
long-range tail of $V_0(r)$ is attractive. Then the quantity $r_*$ is
a measure of the radius of the centrifugal barrier experienced by the
(fermionic) molecules.

At smaller separations, the dipolar interactions between the molecules cause them to depart from the state $|+\rangle$. This occurs when the characteristic interaction energy
$d_c^2/r^3$ becomes larger than the detuning $\hbar|\delta|$, setting the lengthscale $r_\delta=(d_c^2/\hbar\delta)^{1/3}$ entering the matrix (\ref{Hmatrix}). We assume that
$r_{\delta}$ is larger than the confinement length in the $z$-direction, $l_z$, so that the interaction is 2D. The potential curves of even parity are shown in Fig.~2. The
potential $V_0(r)$, being attractive at $r\agt r_{\delta}$ with a long-range $1/r^3$ tail, has a repulsive core for $r\alt r_{\delta}$. The repulsive core prevents low-energy
molecules from approaching each other at distances $r\alt r_{\delta}$ and suppresses inelastic collisions, including ultracold chemical reactions observed at JILA in experiments
with KRb molecules \cite{Ospelkaus2010}. Note that in this respect the MW-dressing of polar molecules in the 2D
geometry can be used for their evaporative cooling.   
 
Actually, the potential curves of even parity, in particular $V_0(r)$,
may cross with potentials of odd parity. However, inelastic
transitions between the states of different parity are not possible in
two-body collisions, and three-body collisions accompanied by such
transitions will be suppressed at the low densities that we consider.

\section{Elastic scattering}

We now consider elastic scattering of fermionic molecules (each in the state $|+\rangle$) undergoing 2D translational motion and interacting with each other via the potential $V_0
(r)$, with the attractive 
dipole-dipole tail (\ref{Vefftail}). At ultralow energies the leading scattering is in the $p$-wave scattering channel. For the investigation of superfluid pairing we need to know
the
off-shell
scattering amplitude defined as
\begin{equation}           \label{offshellampl}
f({\bf k}',{\bf k})=\int \exp(-i{\bf k'\cdot r})V_0(r)\tilde\psi_{{\bf k}}({\bf r})d^2r,
\end{equation}
where $\tilde\psi_{{\bf k}}({\bf r})$ is the true wavefunction of the
relative motion with momentum ${\bf k}$. The $p$-wave part of $f({\bf
  k}',{\bf k})$ is $f(k',k)\exp i\phi$, where
$\phi$ is the angle between the vectors ${\bf k}$ and ${\bf
  k}'$, and the partial $p$-wave amplitude is given by
\begin{equation}          \label{offshellp}
f(k',k)=\int_0^{\infty}J_1(k'r)V_0(r)\tilde\psi(k,r)2\pi rdr,
\end{equation}
with $J_1$ being the Bessel function. The wavefunction of the relative
$p$-wave motion, $\tilde\psi(k,r)$, is governed by the Schr{\" o}dinger equation
\begin{equation}                  \label{Schr1}
-\frac{\hbar^2}{m}\left(\frac{d^2\tilde\psi}{dr^2}+\frac{1}{r}\frac{d\tilde\psi}{dr}-\frac{\tilde\psi}{r^2}\right)+V_0\tilde\psi=\frac{\hbar^2k^2}{m}\tilde\psi.
\end{equation}  
The full on-shell scattering amplitude is obtained from equation (\ref{offshellampl}) at $|{\bf k}'|=|{\bf k}|$, and its $p$-wave part follows from Eq.~(\ref{offshellp}) at $k'=k$,
$f(k)\equiv f(k,k)$. It is related to the $p$-wave scattering phase shift $\delta(k)$ by \cite{LL3}
\begin{equation}           \label{delta}
f(k)\!=\!-\frac{2\hbar^2}{mi}[\exp(2i\delta(k))\!-1]\!=\!-\frac{4\hbar^2}{m}\frac{\tan\delta(k)}{1\!-\!i\tan\delta(k)}.\!
\end{equation}

In Eq.~(\ref{offshellp}) the wavefunction $\tilde\psi(k,r)$ is normalized such that for $r\rightarrow\infty$ we have
$$\tilde\psi(k,r)=J_1(kr)-\frac{if(k)}{4}H_1(kr),$$ 
where $H_1=J_1+iN_1$ is the Hankel function, and $N_1$ is the Neumann function. It is, however, more convenient to normalize the radial wavefunction in such a way that it is real
and for 
$r\rightarrow\infty$ one has 
\begin{equation}         \label{psiass}
\!\!\psi(\!k,\!r\!)\!=\![J_1(kr)\!-\!\tan\delta(k)N_1(kr)]\!\propto\!\cos(kr\!-\!3\pi/4\!+\!\delta).\!\!
\end{equation}
One easily checks that $\tilde\psi(k,r)=\psi(k,r)/(1-i\tan\delta(k))$. Using this relation in Eq.~(\ref{offshellp}) we see that the off-shell scattering amplitude can be written in
the form:
\begin{equation}         \label{fbar}
f(k',k)=\frac{{\bar f}(k',k)}{1-i\tan\delta(k)},
\end{equation}
where ${\bar f}(k',k)$ is real and is given by Eq.~(\ref{offshellp}) with $\tilde\psi(k,r)$ replaced by $\psi(k,r)$. For $k'=k$ one has 
\begin{equation}     \label{fbarkk}
{\bar f}(k,k)\equiv{\bar f}(k)=-(4\hbar^2/m)\tan\delta(k).
\end{equation}

To find the $p$-wave scattering amplitude in the limit $kr_*\ll
1$ we divide the range of distances into two parts: $r<r_0$ and
$r>r_0$, where $r_0$ lies in the interval $r_*\ll r_0\ll k^{-1}$ (see
Fig.~3). As we will see, this separation reflects the presence of two
contributions to the scattering amplitude, the short-range
contribution ($r\alt r_*$) and the anomalous contribution ($r\sim
k^{-1}$).

\begin{figure}[ttp]
\includegraphics[width=0.98\columnwidth]{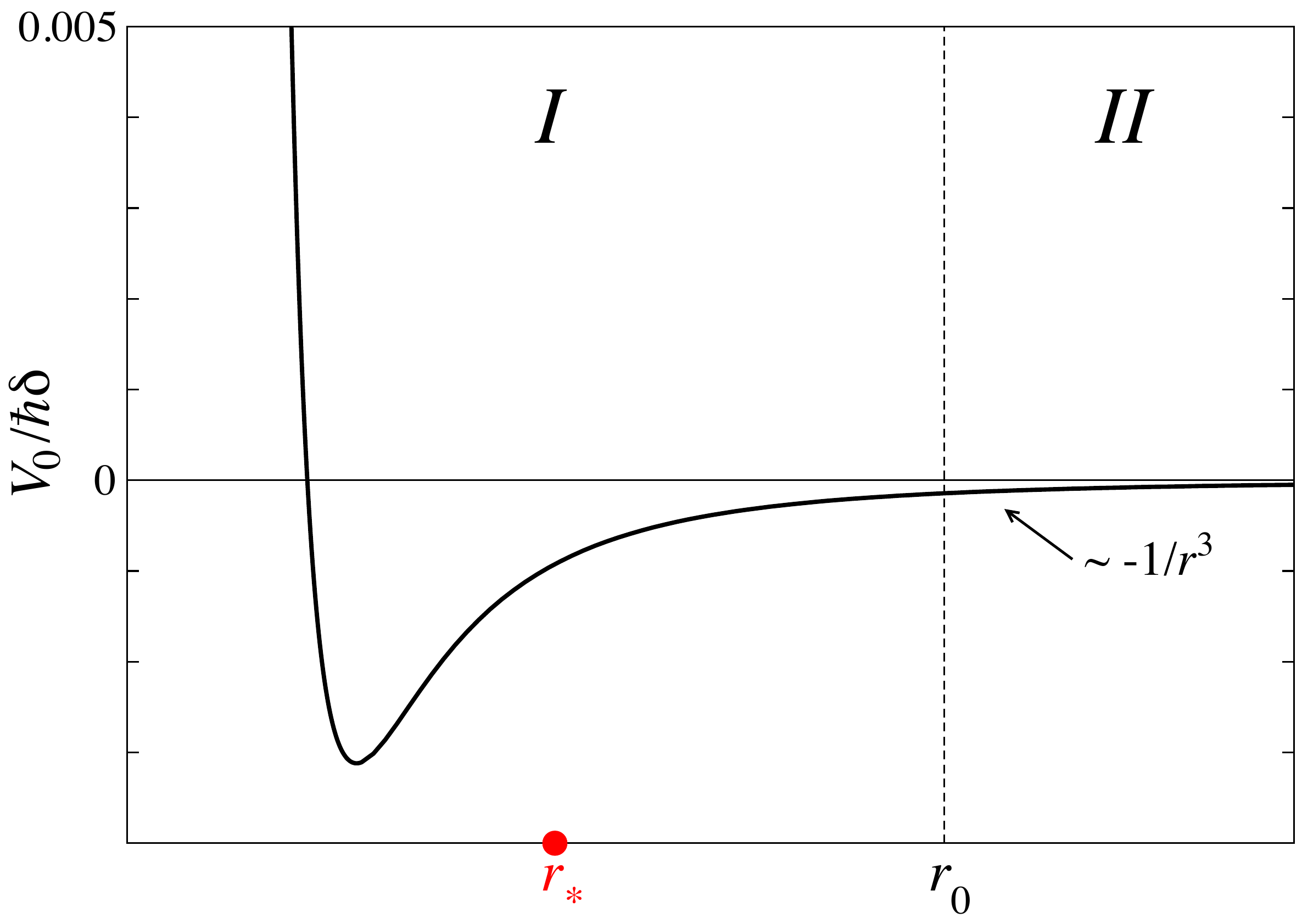}
\caption{ \label{fig:potregions} (color online). The division of ranges
  into regions $I$ $(r<r_0)$ and $II$ $(r>r_0)$. The length scales
  $r_*$ and $r_0\gg r_*$ are indicated by a circle and a dashed line
  respectively.}
\end{figure}

In region I where $r<r_0$, the $p$-wave relative motion of two
fermionic molecules is governed by the Schr{\"o}dinger equation with zero kinetic energy:
\begin{equation}                  \label{Schr2}
-\frac{\hbar^2}{m}\left(\frac{d^2\psi_I}{dr^2}+\frac{1}{r}\frac{d\psi_I}{dr}-\frac{\psi_I}{r^2}\right)+V_0\psi_I=0.
\end{equation}
At distances where‌ the interaction potential is behaving as
$V_0(r)=-\hbar^2r_*/mr^3$, the solution of Eq.~(\ref{Schr2}) reads:
\begin{equation}   \label{psi1}
\psi_I(r)\propto\left[AJ_2\left(2\sqrt{\frac{r_*}{r}}\right)+N_2\left(2\sqrt{\frac{r_*}{r}}\right)\right],
\end{equation}
where the constant $A$ is determined by the behavior of $V_0(r)$ at shorter distances.

In region II, at $r>r_0$, the relative motion is practically free and the potential $V_0(r)$ can be considered a perturbation. To zero order we then have
for the relative wavefunction:
\begin{equation}        \label{psi2}
\psi_{II}^{(0)} (r)=J_1(kr)-\tan\delta_I(k)N_1(kr),
\end{equation}
where the scattering phase shift $\delta_I$ is due to the interaction
between particles in region I. Matching the logarithmic derivatives
of $\psi_I$ and $\psi_{II}^{(0)}$ at $r=r_0$ we obtain
\begin{equation}
\!\!\tan\delta_I\!=\!\frac{\pi k^2r_0r_*}{8}
\!\left[\!1\!-\frac{r_*}{r_0}\!\left(\!2C\!-\frac{1}{2}+\pi A\!-\ln{\frac{r_0}{r_*}}\right)\right]\!, \!\label{deltaI}
\end{equation}
where $C=0.5772$ is the Euler constant, and we took into account that $r_0\gg r_*$ and $kr_0\ll 1$.

We now include perturbatively the contribution to the $p$-wave scattering phase shift from distances $r>r_0$ (region II). In this region, to first order in $V_0(r)$ the relative
wavefunction is given by
\begin{equation}            \label{psi21}
\!\!\psi_{II}^{(1)}(r)\!=\!\psi_{II}^{(0)} (r)-\!\!\int_{r_0}^{\infty}\!\!\!G(r,r')V_0(r')\psi_{II}^{(0)} (r')2\pi r'dr'\!,\!\!
\end{equation}
where the Green function for the free $p$-wave motion obeys the radial equation:
\begin{equation}          \label{Gradial}
-\frac{\hbar^2}{m}\left(\frac{d^2}{dr^2}\!+\!\frac{1}{r}\frac{d}{dr}\!-\!\frac{1}{r^2}+k^2\right)G(r,r')\!=\!\frac{\delta(r-r')}{2\pi r}.
\end{equation}
For the asymptotic representation of the relative wavefunction, chosen in Eq.~(\ref{psiass}), we have
\begin{equation}   \label{G}
G(r,r')=-\frac{m}{4\hbar^2}\left\{\begin{array}{cc}
\psi_{II}^{(0)} (r')N_1(kr), & r>r'\\
\psi_{II}^{(0)} (r)N_1(kr'), & r<r'
\end{array}\right. .
\end{equation}
Then, substituting the Green function (\ref{G}) into equation (\ref{psi21}) and taking the limit $r\rightarrow\infty$ we obtain:
\begin{equation}            \label{delta1eq}
\!\!\!\tan\delta^{(1)}\!(k)\!=\!\tan\delta_I(k)\!-\!\frac{m}{4\hbar^2}\!\!\int_{r_0}^{\infty}\!\!
\left[\psi_{II}^{(0)}(r)\right]^2\!\!V_0(r)2\pi rdr\!. \!\!
\end{equation}
Using Eq.~(\ref{deltaI}) for $\tan\delta_I$ and calculating the integral in Eq.~(\ref{delta1eq}) we find:
\begin{equation}          \label{delta1}
\tan\delta^{(1)}(k)\!=\!\frac{2}{3}kr_*\!-\!\frac{\pi (kr_*\!)^2}{8}\!\left[\ln{\frac{r_*}{r_0}}\!+\!2C\!-\!\frac{3}{2}\!+\!\pi A\right],
\end{equation}
and we omitted terms that contain higher powers of $k$.

Adding the second order contribution we have for the relative wavefunction:
\begin{eqnarray}         
\psi_{II}^{(2)}(r)=&&\psi_{II}^{(1)}(r)+\int_{r_0}^{\infty}G(r,r')V_0(r')2\pi r'dr' \nonumber \\
&&\hspace{-6mm}\times \int_{r_0}^{\infty}G(r',r'')V_0(r'')\psi_{II}^{(0)} (r'')2\pi r''dr''.\label{psi22}
\end{eqnarray}
  
Then, using Eq.~(\ref{G}) and taking the limit $r\rightarrow\infty$ we
see that including the second order contribution, the scattering phase
shift becomes
\begin{eqnarray}         \label{deltafin}
\tan\delta(k)=&&\tan\delta^{(1)}(k)-\frac{m^2}{8\hbar^4}\int_{r_0}^{\infty}\left[\psi_{II}^{(0)}
  (r)\right]^2
V_0(r)2\pi rdr \nonumber \\
&&\times\int_r^{\infty}N_1(kr')V_0(r')\psi_{II}^{(0)} (r')2\pi r'dr'.
\end{eqnarray}
As we are not interested in terms that are proportional to $k^3$ or
higher powers of $k$ and in terms that behave as $(kr_*)^2r_*/r_0$, we
may omit the term $\tan\delta_I(k)N_1 (kr)$ in the expression for
$\psi_{II}^{(0)}(r)$. Then equation (\ref{deltafin}) reduces to
\begin{eqnarray}
&&\hspace{-3mm}\tan\delta(k)=\tan\delta^{(1)}(k)-\frac{(\pi kr_*)^2}{2}\int_{kr_0}^{\infty}\frac{J_1^2(x)}{x^2}dx\times \nonumber \\
&&\Big[\frac{2}{3}x(N_0(x)J_2(x)-N_1(x)J_1(x))-\frac{1}{2}N_0(x)J_1(x) \nonumber \\
&&\,\,+\frac{1}{6}N_1(x)J_2(x)-\frac{1}{\pi x}\Big]. 
\end{eqnarray}
For the first four terms in square brackets we may put the lower limit
of integration equal to zero. Then, using the relations:
\begin{eqnarray}
&&\int_0^{\infty}J_1^3(x)N_1(x)\frac{dx}{x}=-\frac{1}{4\pi} ,  \nonumber \\
&&\int_0^{\infty}J_1^2(x)J_2(x)N_0(x)\frac{dx}{x}=\frac{1}{8\pi}, \nonumber \\
&&\int_0^{\infty}J_1^3(x)N_0(x)\frac{dx}{x^2}=\frac{1}{16\pi}, \nonumber \\
&&\int_0^{\infty}J_1^2(x)J_2(x)N_1(x)\frac{dx}{x^2}=-\frac{1}{16\pi}, \nonumber \\
&&\int_{kr_0}^{\infty}J_1^2(x)\frac{dx}{x^3}\approx \frac{1}{16}-\frac{C}{4}+\frac{\ln{2}}{4}-\frac{1}{4}\ln{kr_0}, \nonumber 
\end{eqnarray}
we obtain
\begin{eqnarray}         
\!\tan\delta(k)\!=\tan\delta^{(1)}\!(k)\!-\!\frac{\pi(kr_*)^2}{8}\!\left\{\!\frac{7}{12}\!+\!C\!-\!\ln{2}\!+\!\ln{kr_0}\!\right\}. \nonumber
\end{eqnarray}
Substituting $\tan\delta^{(1)}(k)$ from Eq.~(\ref{delta1}) we eventually arrive at the scattering phase shift
\begin{equation}          \label{deltafinal}
\tan\delta(k)=\frac{2}{3}kr_*-\frac{\pi(kr_*)^2}{8}\ln{\rho kr_*},
\end{equation} 
where $\rho=\exp\{3C-\ln{2}-11/12+\pi A\}\simeq 1.13\exp(\pi A)$.

Using equations (\ref{delta}) and (\ref{fbarkk}) we then immediately obtain the on-shell scattering amplitude $f(k)$ and the amplitude ${\bar f}(k)$. Note 
that ${\bar f}(k)$ is conveniently represented as a sum of two terms: ${\bar f}(k)={\bar f}_1(k)+{\bar f}_2(k)$, where
\begin{eqnarray}
&&{\bar f}_1(k)=-\frac{8}{3}\frac{\hbar^2}{m}kr_*,   \label{f1k}  \\
&&{\bar f}_2(k)=\frac{\pi }{2}\frac{\hbar^2}{m}(kr_*)^2\ln{\rho kr_*},  \label{f2k}
\end{eqnarray}
so that in the low-momentum limit the term ${\bar f}_1(k)$ is dominant. The related tangent of the scattering phase shift is contained in the first order contribution from
distances
$r>r_0$, given by the second term on the right-hand side of Eq.~(\ref{delta1eq}) in which one keeps only $J_1(kr)$ in the expression for $\psi_{II}(r)$. This says that 
${\bar f}_1(k)=-(4\hbar^2/m)\tan\delta(k)=\int_{r_0}^{\infty}J_1^2(kr)V_0(r)2\pi rdr$ and for $V_0$ given by equation (\ref{Vefftail}) the amplitude ${\bar f}_1(k)$ is reduced
exactly to the
result of Eq.~(\ref{f1k}).

The off-shell scattering amplitude ${\bar f}(k',k)$ defined by Eq.~(\ref{fbar}) can also be written as ${\bar f}_1(k',k)+{\bar f}_2(k',k)$, and the leading low-momentum
contribution
is given by 
\begin{eqnarray}        
{\bar f}_1(k',k)&&=\int_{r_0}^{\infty}J_1(k'r)J_1(kr)V_0(r)2\pi rdr  \nonumber \\
&&=-\frac{\pi\hbar^2}{m} kr_*F\left(-\frac{1}{2},\frac{1}{2},2,\frac{k^2}{k'^2}\right), \label{f1kk}
\end{eqnarray}
where we took into account that for $r>r_0$ the interaction potential has the form $V_0(r)=-\hbar^2r_*/mr^3$ . The quantity $F$ in Eq.~(\ref{f1kk}) is the hypergeometric
function, 
and the result is written for $k<k'$. For $k>k'$ one should interchange $k$ and $k'$.     

\section{Inelastic collisional processes}

For the considered case of $\delta>0$ the dressed molecular state
$|+\rangle$ is higher in energy than the states $|-\rangle$ and
$|\Phi_{1-1}\rangle$. Therefore, molecules in the dressed state $
|+\rangle$ may undergo pair inelastic collisions in which one or both are transferred to the state $|-\rangle$ or $|\Phi_{1-1}\rangle$.   For  $\Omega_R\alt\delta$, 
the released kinetic energy is $\sim\hbar\delta$ and it can cause both molecules to escape from the sample. The kinetic energy release requires a momentum transfer of
$\sim \hbar/\lambda_\delta$ with $\lambda_\delta\equiv\sqrt{\hbar/m\delta}$.  For ${\lambda_\delta}/{r_\delta} \ll 1$ the particles cannot approach each
other sufficiently closely to allow the required momentum exchange, and one anticipates a reduction in the loss rate. The same condition can be derived semiclassically as the
condition of adiabatic motion in the potential $V_0(r)$.

In order to go beyond this limit and determine the loss rate for
$\lambda_{\delta}$ approaching $r_{\delta}$, we have solved the full
two-body scattering problem, involving states of even parity which at
an infinite separation are:
$(|+\rangle,|+\rangle),\,(|+\rangle,|-\rangle),\,(|+\rangle,
|\Phi_{1-1}\rangle),\,(|-\rangle,|-\rangle),$ and
$(|-\rangle,|\Phi_{1-1}\rangle)$ [the state
$(|\Phi_{1-1}\rangle,|\Phi_{1-1}\rangle)$ is decoupled].  We calculate
numerically the probabilities $P_l$ that two $|+\rangle$-state
molecules with relative angular momentum $l$ are scattered into {\it
  any} outgoing channel in which at least one of them is in the state
$|-\rangle$ or $|1,-1\rangle$. This corresponds to non-adiabatic
transitions from the potential $V_0(r)$ to the other potentials shown
in the inset of Fig.~2. The Hamiltonian term which causes these
non-adiabatic transitions is the kinetic energy (Laplacian)
term. Defining the spinor $\chi_i(r)=\sqrt{r}\psi_i(r)$, where the
index $i$ labels the two-particles states
$(|+\rangle,|+\rangle),\,(|+\rangle,|-\rangle),\,(|+\rangle,|\Phi_{1-1}\rangle),\
(|-\rangle,|-\rangle),$ and $(|-\rangle,|\Phi_{1-1}\rangle)$, we
obtain that the Laplacian term acts on $\chi_i(r)$ as
$$\left[\hat L\chi(r)\right]_i=L_{ij}(r)\chi_j(r).$$
The matrix $L_{ij}(r)$ is diagonal, $L_{ij}(r)={\bar L}_i\delta_{ij}$, with
\begin{eqnarray} 
\!\!&&\!\!\!\!{\bar L}_i(r)=\frac{\hbar^2}{m}\left(-\frac{d^2}{dr^2}+\frac{l^2-1/4}{r^2}\right);\,\,i=1,2,4, \label{Lbar125} \\
\!\!&&\!\!\!\!{\bar L}_i(r)=\frac{\hbar^2}{m}\left(-\frac{d^2}{dr^2}+\frac{(l-2)^2-1/4}{r^2}\right);\,\,i=3,5. \label{Lbar35} 
\end{eqnarray}
We note that coupling between these scattering channels will affect
also the {\it elastic} scattering of two molecules incident in the
$i=1$ channel, in such a way that the scattering amplitude will differ
for molecules in the $l = +1$ and $l =-1$ channels. This difference
is a consequence of the fact that time reversal symmetry is broken by
the circularly polarized MW field.  The change in the scattering
amplitude is small for $r_\delta \ll r^*$, which is valid in the
situations studied below, so it will be a small effect.  Nevertheless,
we note that it will cause a slight difference in the energetics (and
the transition temperatures) of the $p_x+ ip_y$ and $p_x- ip_y$
phases, providing a small symmetry breaking perturbation that will
favour one of these two phases. Which one is favoured depends on the
handedness of the circularly polarized MW field.

Taking into account that two molecules are lost in each inelastic collision, and writing the molecule loss rate as 
\begin{equation}           \label{ndot}
\dot n=-\alpha n^2, 
\end{equation}
for the 2D inelastic rate constant we obtain:
\begin{equation}     \label{alpha}
\alpha = 4{\hbar}/{m}\sum_{l}P_{l}. 
\end{equation}
This rate constant at zero static electric field $E_{dc}$ has been calculated in Ref.~\cite{Cooper2009}. As in \cite{Cooper2009}, we treat particles of the outgoing channel
as 2D. This is surely valid for $\lambda_{\delta}\gg l_z$, where $l_z$ is the size of the molecule wavefunction in the tightly confined $z$-direction (amplitude of
zero point oscillations). In this case the energy release is insufficient to allow transitions to excited states in the $z$-direction. On the other hand, the 2D treatment
of outgoing particles is relevant in the opposite limiting case of $\lambda_{\delta}\ll l_z$ where they have a very small angle $\sim\lambda_{\delta}/l_z$ out of the
2D plane. Thus, this approach should give a good result also for the intermediate case $\lambda_{\delta}\sim l_z$.

The dependence of $\alpha$ on $r_{\delta}/\lambda_{\delta}$ at a constant $kr_*$, obtained in Ref.~\cite{Cooper2009} for $\beta=0$, shows the general trend of a reduction  
of inelastic losses with increasing $r_\delta/\lambda_\delta$, which is consistent with the semiclassical expectations. In addition, there is a dramatic modulation of
the inelastic scattering rate, arising from an interference of incoming and outgoing waves in the scattering potential. We recover these results and investigate the 
dependence of $\alpha$ on $kr_*$. Fig.~4 shows $\alpha$ as a function of $kr_*$ for $\Omega_R/\delta=0.25$, $r_{\delta}/\lambda_{\delta}=8$, and $\beta=0$. The dependence
$\alpha\propto (kr_*)^2$ expected in the low-momentum limit is observed for $kr_*\alt 0.1$. 

\begin{figure}[ttp]
\includegraphics[width=.98\columnwidth]{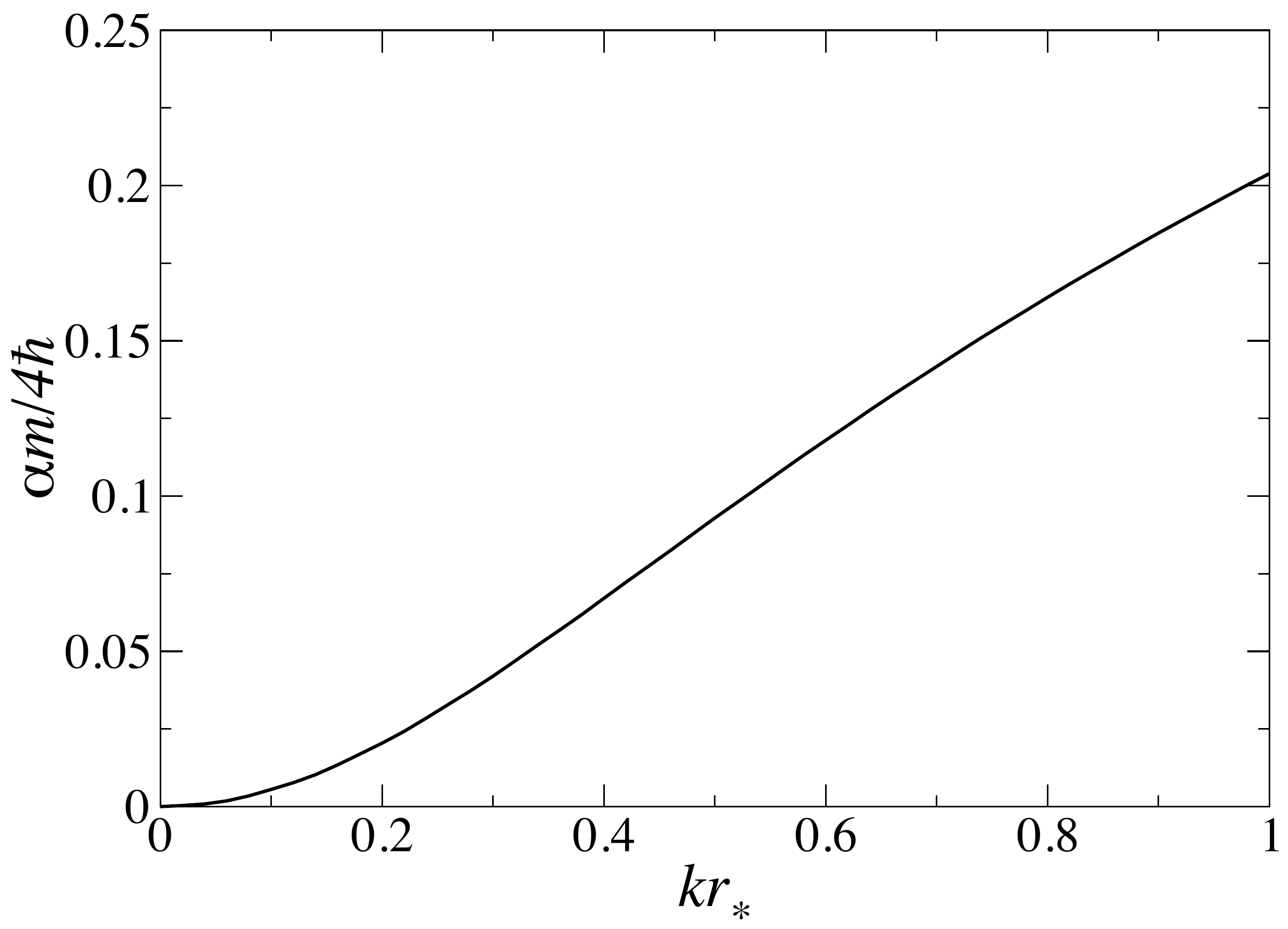}
\caption{ \label{fig:alphavskrstar} 
Inelastic rate constant $\alpha$ as a function of $kr_*$ for
$\Omega_R=0.25\delta$, $r_\delta/\lambda_\delta=8$, and $\beta=0$.
}
\end{figure}

We also consider the influence of a static electric field on the inelastic losses. In Fig.~5 we present the inelastic rate constant $\alpha$ versus the ratio
$r_{\delta}/\lambda_{\delta}$ for various values of $\beta$ at $kr_*=1$. One sees that an increase in $\beta$ shifts the interference minimum of $\alpha$ towards larger values of
$r_{\delta}/\lambda_{\delta}$ and leads to lower values of $\alpha$ at these points. For example, for $\beta=0.2$ the (first) interference minimum is located at
$r_{\delta}/\lambda_{\delta}=15.3$ with $\alpha\simeq 4\times 10^{-5}\hbar/m$, whereas for $\beta=0$ the minimum is at $r_{\delta}/\lambda_{\delta}=10.7$ and the corresponding
inelastic rate constant is $\alpha\simeq 3\times 10^{-4}\hbar/m$.

\begin{figure}[ttp]
\includegraphics[width=1.\columnwidth]{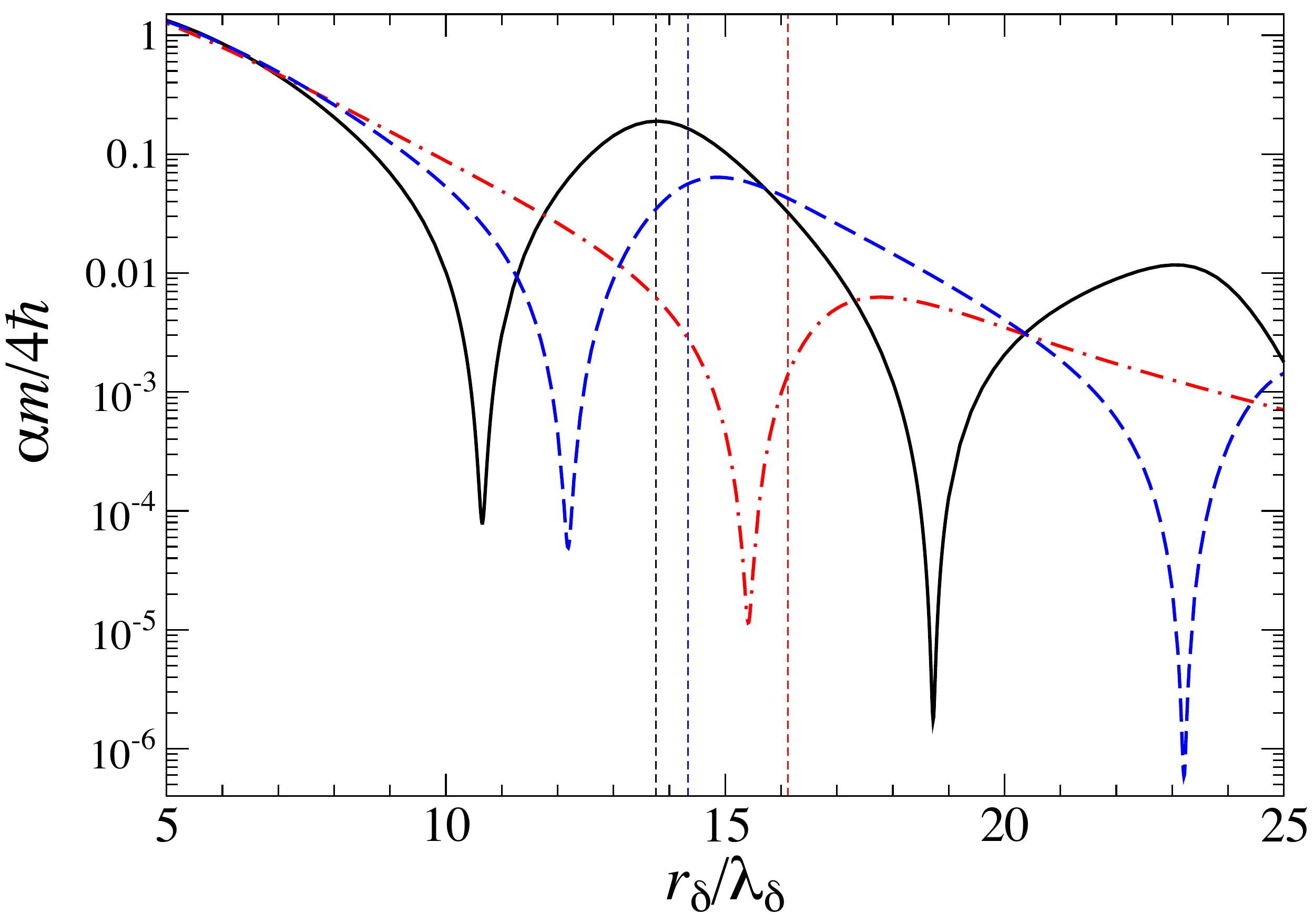}
\caption{
(color online). Inelastic rate constant $\alpha$ as a function of
$r_\delta/\lambda_\delta$ for $\Omega_R=0.25\delta$, $kr_*=1$, and $\beta=0$
(black, solid), $\beta=0.1$ (blue, dashed), and $\beta=0.2$ (red, dash-dotted). The
critical value of $r_\delta/\lambda_\delta$ above which a bound state
of two molecules exists is indicated with corresponding vertical lines.
\label{fig:rates}}
\end{figure}

The characteristic lifetime of the sample is $\tau\sim(\alpha n)^{-1}$, and for a deeply degenerate molecular Fermi gas one should take the rate constant $\alpha$ at $k$ equal to
the Fermi momentum $k_F$. We then see that even for $k_Fr_*=1$ the lifetime $\tau$ of, for example, Na$^{40}$K molecules ranges from $20$ to $2$ s when increasing the density
from $10^8$ to $10^9$ cm$^{-2}$ and considering the first interference minimum at $\beta=0.2$ (see Fig.~5). 

We should avoid the presence of bound states of two molecules in the potential $V_0(r)$. Otherwise three-body recombination will lead to a rapid decay of the gas \cite{Cooper2009}. 
A dimensional estimate for the three-body recombination rate constant gives $\alpha_{{\rm rec}}\sim (\hbar r^{*2}/m)(k_Fr^*)^4$.  The corresponding decay time is $\tau_{{\rm
rec}}\sim (\alpha_{{\rm rec}}n^2)^{-1}$, and it can be as short as milliseconds or even lower at densities $n\sim 10^8 - 10^9$ cm$^{-2}$ for not very small values of $k_Fr_*$. As
we will see below, an efficient superfluid pairing and sufficiently high superfluid transition temperature require $k_Fr_*$ approaching unity.

Bound molecule-molecule states in the potential $V_0(r)$ appear under an increase in $r_{\delta}/\lambda_{\delta}$. Critical values of $r_{\delta}/\lambda_{\delta}$ below which
bound states are absent for a given $\beta$, are shown by dashed vertical lines in Fig.~5. Thus, to the left of these lines the three-body recombination is absent. This is in
particular the case at the first interference minimum for which the lifetime of the gas due to two-body inelastic collisions was estimated above.

\section{Superfluid pairing and the gap equation} 

We now discuss superfluid pairing in the 2D gas of MW-dressed
fermionic polar molecules in the internal state $|+\rangle$,
interacting with each other via the potential $V_0(r)$.  The
Hamiltonian of the system reads:
\begin{eqnarray}         
&&\hspace{-10mm}
\hat{\cal H}=\int d^2r\,\hat\psi^{\dagger}({\bf r})\left(-\frac{\hbar^2}{2m}\nabla^2-\mu\right)\hat\psi({\bf r}) \nonumber  \\
&&\hspace{-4mm}
+\frac{1}{2}\int d^2rd^2r'\hat\psi^{\dagger}({\bf r})\hat\psi^{\dagger}({\bf r}')V_0({\bf r}-{\bf r}')\hat\psi({\bf r}')\hat\psi({\bf r}),  \label{calH}  
\end{eqnarray}
where $\hat\psi({\bf r})$ is the field operator of the dressed molecules, and $\mu$ is the chemical potential. The pairing is due to an effective attractive interaction between the
fermions, which is reflected by a negative sign of the scattering
amplitude ${\bar f}(k)$ at $k$ close to the Fermi momentum $k_F$ as
follows from Eqs. (\ref{f1k}) and
(\ref{f2k}). We thus employ a BCS approach \cite{LL9,Leggett} and reduce $\hat{\cal H}$ to a bilinear form:
\begin{eqnarray}
&&\hspace{-10mm}
\hat{\cal H}_{BCS}=\int d^2r\,\hat\psi^{\dagger}({\bf r})\left(-\frac{\hbar^2}{2m}\nabla^2-\mu\right)\hat\psi({\bf r})  \nonumber  \\ 
&&\hspace{-2mm}+
\frac{1}{2}\int d^2rd^2r'[\Delta^*({\bf r},{\bf r}')\hat\psi({\bf
  r}')\hat\psi({\bf r})+h.c.].
\label{HBCS}
\end{eqnarray}
The order parameter (gap) obeys the gap equation
\begin{equation}          \label{gapcoord}
\Delta({\bf r},{\bf r}')=V_0({\bf r}-{\bf r}')\langle\hat\psi({\bf r}')\hat\psi({\bf r})\rangle,
\end{equation}
with the symbol $\langle\,\rangle$ denoting the expectation value. Strictly speaking, in two dimensions this approach is accurate at $T=0$. At non-zero temperatures the long-range
order is destroyed by long-wave thermal fluctuations of the phase and one has only an algebraic order. However, in the weakly interacting regime the characteristic phase coherence
length is exponentially large, and on a distance scale smaller than this length one may still use the order parameter (\ref{gapcoord}) and the BCS approach. In the uniform case
this order parameter depends on ${\bf r}$ and ${\bf r}'$ only through the difference $({\bf r}-{\bf r}')$. 

The Hamiltonian $\hat{\cal H}_{BCS}$ is reduced to a diagonal form $\hat{\cal H}_{BCS}=\sum_{{\bf k}}\epsilon_k\hat b^{\dagger}_{{\bf k}}\hat b_{{\bf k}}+{\rm const}$, by using the
Bogoliubov transformation
$$\hat\psi({\bf r})=\frac{1}{\sqrt{S}}\sum_{{\bf k}}[u_{{\bf k}}\exp(i{\bf kr})\hat b_{{\bf k}}+v_{{\bf k}}^*\exp(-i{\bf kr})\hat b^{\dagger}_{{\bf k}}].$$
Here $S$ is the surface area of the system, while $\hat b_{{\bf
    k}},\hat b^{\dagger}_{{\bf k}}$ and
$\epsilon_k=\sqrt{\xi_k^2+|\Delta_{{\bf k}}|^2}$ are
annihilation/creation operators and energies of single-particle
excitations. The functions $u_{{\bf k}}$ and $v_{{\bf k}}$ satisfy the
well-known Bogoliubov-de Gennes equations. They are given by
$$u_{{\bf k}}=\frac{\xi_k+\epsilon_k}{\sqrt{2\epsilon_k(\xi_k+\epsilon_k)}};\,\,\,\,\,v_{{\bf k}}=\frac{\Delta_{{\bf k}}}{\sqrt{2\epsilon_k(\xi_k+\epsilon_k)}},$$    
where $\Delta_{{\bf k}}=\int d^3r\Delta({\bf r}-{\bf r}')\exp[i{\bf k}({\bf r}-{\bf r}')]$ is the momentum-space order parameter, $\xi_k=\hbar^2k^2/2m-\mu$, and the chemical
potential in the weakly interacting regime is close to the Fermi energy $E_F=\hbar^2k_F^2/2m$. In the momentum space, the gap equation (\ref{gapcoord}) takes the form: 
\begin{equation}    \label{momentumgapequation}
\Delta_{{\bf k}}=-\int\frac{d^2q}{(2\pi)^2}V_0({\bf q}-{\bf k})\Delta_{{\bf q}}{\cal K}(q),
\end{equation}
where $V_0({\bf q})$ is the Fourier transform of the potential $V_0(r)$, and ${\cal K}(q)=\tanh(\epsilon_q/2T)/2\epsilon_q$.  

We now renormalize the gap equation (\ref{momentumgapequation}) by expressing the Fourier transform of $V_0(r)$ through the off-shell scattering amplitude. The relation between
these two
quantities reads \cite{LL3}:
\begin{equation}           \label{fV}
f({\bf k}',{\bf k})=V_0({\bf k}'-{\bf k})+\int\frac{d^2q}{(2\pi)^2}\,\frac{V_0({\bf k}'-{\bf q})f({\bf q},{\bf k})}{2(E_k-E_q-i0)}.
\end{equation}
Multiplying Eq.~(\ref{fV}) by ${\cal K}(k')\Delta_{{\bf k}'}$ and integrating over $d^2k'$, with the help of Eq.~(\ref{momentumgapequation}) we then obtain:
\begin{equation}          \label{renormgapequation}
\!\!\!\Delta_{{\bf k}}\!=\!-\!\!\!\int \!\! \!f({\bf k}'\!,{\bf k})\Delta_{{\bf k}'}\!\!\left[{\cal K}(k')\!-\!\frac{1}{2(E_{k'}\!-\!E_k\!-\!i0)}\right]\!\frac{d^2k'}{(2\pi)^2}.\!
\end{equation}  

Note that in contrast to the commonly used renormalization procedure expressing the Fourier transform of the interaction potential through the vertex function $\Gamma({\bf k},{\bf
k}',E)$ \cite{AGD} at an arbitrarily chosen energy $E$, which coincides with $f({\bf k}',{\bf k})$ for $E=E_{k'}$, we use the off-shell scattering amplitude from the very
beginning of the renormalization.

At $T=0$ we put $\tanh(\epsilon_k/2T)=1$ and, hence, ${\cal
  K}(k)=1/2\epsilon_k$. We then perform an analysis assuming that in
the weak coupling limit the main contribution to the integral in
Eq.~(\ref{renormgapequation}) comes from momenta $k'$ close to
$k_F$. It shows that the dominant pairing instability is in the
channel with orbital angular momentum $l=1$, since for higher angular
momenta the interaction (scattering) amplitude is much smaller. The
most stable low temperature phase has $p_x\pm i p_y$ symmetry,
following from the fact that this phase fully gaps the Fermi surface,
in contrast to competing phases \cite{Andersonmorel}. A full numerical
solution of the regularized gap equation confirms this analysis.

\begin{figure}[ttp]
\includegraphics[width=.95\columnwidth]{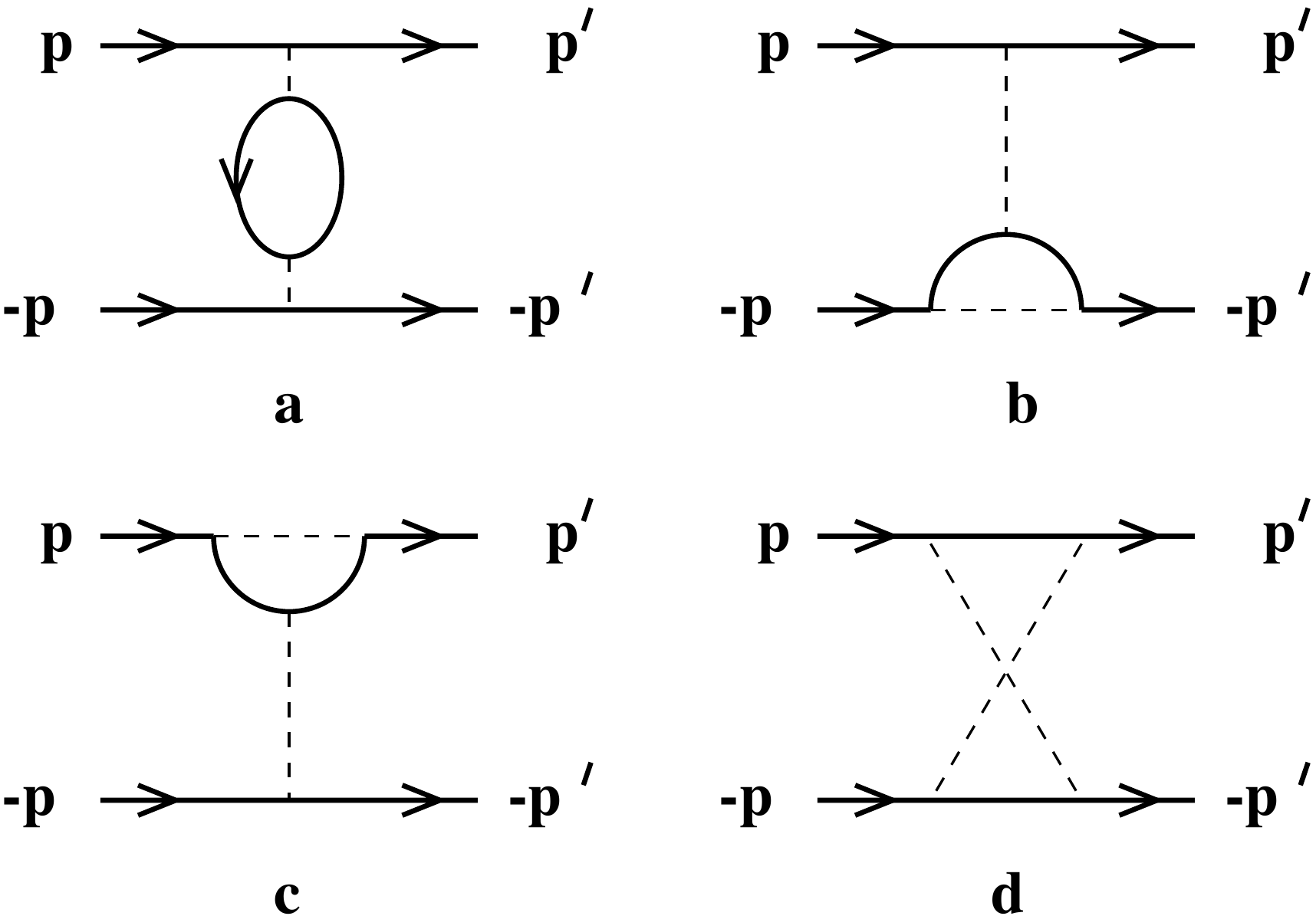}
\caption{
The lowest order many-body corrections to the effective interparticle
interaction.
\label{fig:gm}}
\end{figure}

In fact, equation (\ref{momentumgapequation}) and, hence, equation
(\ref{renormgapequation}) are not sufficient for obtaining a correct
result for the order parameter. One should calculate the quantity $\delta
V({\bf q},{\bf k})$ originating from many-body effects and add it to $V_0({\bf
  q}-{\bf k})$ in Eq.~(\ref{momentumgapequation}). The quantity  $\delta V({\bf q},{\bf k})$ is a
correction to the bare interparticle interaction $V_0$, and the
leading terms of $\delta V({\bf q},{\bf k})$ are second order in $V_0$
\cite{GM}. The corresponding diagrams are shown in Fig.~\ref{fig:gm}
and they are the same as in the case of superfluid pairing between
identical dipolar fermions in three dimensions
\cite{Baranov2002}. They describe processes in which one of the
two colliding particles polarizes the medium by creating a
particle-hole pair. In Fig. \ref{fig:gm}(a) the particle-hole pair then
annihilates due to the interaction with the other colliding
particle. In \ref{fig:gm}(b), \ref{fig:gm}(c), and \ref{fig:gm}d the hole
annihilates together with one of the colliding particles. In
\ref{fig:gm}(b) and \ref{fig:gm}(c) the particle-hole pair is created due
to the interaction of the medium with one of the colliding particles,
and the hole annihilates with the other colliding partner. In
\ref{fig:gm}(d) these creation and annihilation processes involve one
and the same colliding particle. Including these many-body effects,
which were introduced by Gor'kov and Melik-Barkhudarov \cite{GM}, the
gap equation becomes:
\begin{eqnarray}      
\!\!\!&&\Delta_{{\bf k}}\!=\!-\!\!\int\! f({\bf k}',{\bf k})\Delta_{{\bf k}'}\!\left[{\cal K}(k')-\!\frac{1}{2(E_{k'}-E_k-i0)}\right]\frac{d^2k'}{(2\pi)^2}  \nonumber \\
&&\hspace{8mm}
-\int\delta V({\bf k}',{\bf k}){\cal K}(k')\Delta_{{\bf k}'}\frac{d^2k'}{(2\pi)^2}.  \label{gapGMequation}
\end{eqnarray}

It turns out that we should also take into account the difference between particles and quasiparticles (single-particle excitations). The latter are characterized by the effective
mass $m_*$, and the density of states near the Fermi surface is 
\begin{equation}           \label{nuF}
\nu_F=\frac{m_*}{2\pi\hbar^2}.
\end{equation}
Thus, dealing with this quantity in the gap equation we should replace
$m$ with $m_*$. Note that for short-range potentials the effective
mass correction can be neglected.  However, for fairly long-range
interactions like $1/r^3$ in 2D, this correction becomes important
as was recently demonstrated for the case of $s$-wave pairing in
bilayer dipolar systems \cite{Baranov2011}.

According to the Landau Fermi-liquid theory, the derivative of the quasiparticle energy $\epsilon(k)$ with respect to momentum is given by \cite{LL9}:
\begin{equation}         \label{LFL1}
\frac{\partial\epsilon(k)}{\partial {\bf k}}=\frac{\hbar^2 {\bf k}}{m}+\int F({\bf k},{\bf k}')\frac{\partial N(k')}{\partial{\bf k}'}\frac{d^2k'}{(2\pi)^2},
\end{equation}
where $N(k')$ is the distribution function of quasiparticles for which we take a step function $N(k')=\theta(k_F-k')$, so that $\partial N(k')/\partial{\bf k}'=-({\bf
k}'/k')\delta(k'-k_F)$. The leading contribution to the interaction function of quasiparticles, $F({\bf k},{\bf k}')$, comes from the scattering by the $1/r^3$ tail of the
interaction potential $V_0(r)$. This contribution can be calculated in the first Born approximation and, including all (odd) partial waves, for small momenta it is given by
\begin{equation}         \label{F}
F({\bf k},{\bf k}')=[V_0(0)-V_0({\bf k}-{\bf k}')]=-\frac{2\pi\hbar^2}{m}|{\bf k}-{\bf k}'|r_*.
\end{equation}
This simply follows from the fact that in our weakly interacting system of identical fermionic particles the interaction energy par unit area is
\begin{eqnarray}
&&E_{\rm int}=\frac{1}{2}\sum_{{\bf k}_1,{\bf k}_2,{\bf k}_3}V_0({\bf k}_3-{\bf k}_2)\langle \hat a^{\dagger}_{{\bf k}_1+{\bf k}_2-{\bf k}_3}\hat a^{\dagger}_{{\bf k}_3}
\hat a_{{\bf k}_2}\hat a_{{\bf k}_1}\rangle  \nonumber \\
&&=\frac{1}{2}\sum_{{\bf k}_1,{\bf k}_2}[V_0(0)-V_0({\bf k}_1-{\bf k}_2)]N(k_1)N(k_2), \nonumber
\end{eqnarray}
with $\hat a^{\dagger}$ and $\hat a$ being the (quasi)particle operators.

For $k$ near $k_F$ equation (\ref{LFL1}) then yields:
\begin{equation}          \label{LFL2}
v_F=\frac{\hbar k_F}{m}-\frac{1}{\hbar}\int_0^{2\pi}F(2k_F|\sin{\phi/2}|)\cos{\phi}\frac{k_Fd\phi}{4\pi^2},
\end{equation}
where $\phi$ is the angle between the vectors ${\bf k}$ and ${\bf k}'$, and $v_F=\partial\epsilon/\hbar\partial k|_{k=k_F}$ is the Fermi velocity. The effective mass is defined as
$m_*=\hbar k_F/v_F$. Then, using Eq.~(\ref{F}) we have:
\begin{equation}           \label{meff}
\frac{m_*}{m}=1+\frac{4}{3\pi}k_Fr_*.
\end{equation}  

The Gor'kov-Melik-Barkhudarov corrections and the replacement of the bare mass $m$ by $m_*$ do not change our conclusion that the ground state has $p_x\pm ip_y$ symmetry.

\section{Order parameter and transition temperature}

In the 2D geometry that we consider, the transition temperature $T_c$ of a Fermi gas from the normal to superfluid regime is set by the Kosterlitz-Thouless transition. However, in
the weak coupling limit the Kosterlitz-Thouless temperature is very close to $T_c$ calculated in the BCS approach \cite{Miyake}. The latter follows from Eq.~(\ref{gapGMequation})
as the highest temperature at which this equation has a non-trivial solution for the order parameter. Using the $p_x+ip_y$ symmetry we write $\Delta_{{\bf
k}}=\Delta(k)\exp(i\phi_{{\bf k}})$. Then, multiplying both sides of Eq.~(\ref{gapGMequation}) by $\exp(-i\phi_{{\bf k}})$ and integrating over $d\phi_{{\bf k}}$ and $d\phi_{{\bf
k}'}$ we obtain the same equation (\ref{gapGMequation}) in which $\Delta_{{\bf k}}$ and $\Delta_{{\bf k}'}$ are replaced by $\Delta(k)$ and $\Delta(k')$, the off-shell scattering 
amplitude $f({\bf k}',{\bf k})$ is replaced by its $p$-wave part $f(k',k)$ defined in Eq.~(\ref{offshellp}), and $\delta V({\bf k}',{\bf k})$ by its $p$-wave part $\delta V(k',k)$.
Calculating the contribution of the pole in the second term in the square brackets and expressing the off-shell scattering amplitude through ${\bar f}(k',k)$ by using
Eq.~(\ref{fbar}), we obtain: 
\begin{eqnarray}         
\!\!\!&&\Delta(k)\!=\!-{\rm P}\!\!\int\! \!{\bar f}(k',k)\Delta(k')\!\left[{\cal K}(k')\!-\frac{1}{2(E_{k'}\!-\!E_k)}\right]\!\frac{d^2k'}{(2\pi)^2}   \nonumber \\ 
&&\hspace{11mm}
-\int\delta V(k',k){\cal K}(k')\Delta(k')\frac{d^2k'}{(2\pi)^2}.      \label{gappwave1}
\end{eqnarray}
The symbol ${\rm P}$ denotes the principal value. The amplitude ${\bar
  f}(k',k)$ is real, and Eq. (\ref{gappwave1}) is convenient for
analytical and numerical calculations of the order parameter and
$T_c$.

The leading contribution to the integral on the right hand side of Eq.~(\ref{gappwave1}) is related to the first term in square brackets and comes from a narrow vicinity of the
Fermi surface. Omitting other contributions we then establish a relation between $\Delta(k)$ and the order parameter on the Fermi surface:
\begin{equation}          \label{DeltakkF}
\Delta(k)=\Delta(k_F)\frac{{\bar f}(k_F,k)}{{\bar f}(k_F)}.
\end{equation}  
For small momenta it is sufficient to use Eq.~(\ref{f1kk}) for ${\bar f}(k_F,k)$ and Eq.~(\ref{f1k}) for $\bar f(k_F)$. This yields:
\begin{equation}          \label{DkF}
\Delta(k)\!=\!\frac{3\pi\Delta(k_F)}{8}\!\left\{\begin{array}{cc}
\!\frac{k}{k_F}F\left(-\frac{1}{2},\frac{1}{2},2,\frac{k^2}{k_F^2}\right)\!, & k<k_F\\
F\left(-\frac{1}{2},\frac{1}{2},2,\frac{k_F^2}{k^2}\right)\!, & k>k_F
\end{array}\right.
\end{equation}
so that $\Delta(k)\propto k$ for $k<k_F$, and it becomes $k$-independent for $k>k_F$. In Fig.~7 we compare the result of Eq.~(\ref{DkF}) with the result of full scale numerics at
$T=0$ for the interaction potential $V_0(r)$ at $\Omega_R=0.25\delta$ and $\beta=0$. 

\begin{figure}[ttp]
\includegraphics[width=1.08\columnwidth]{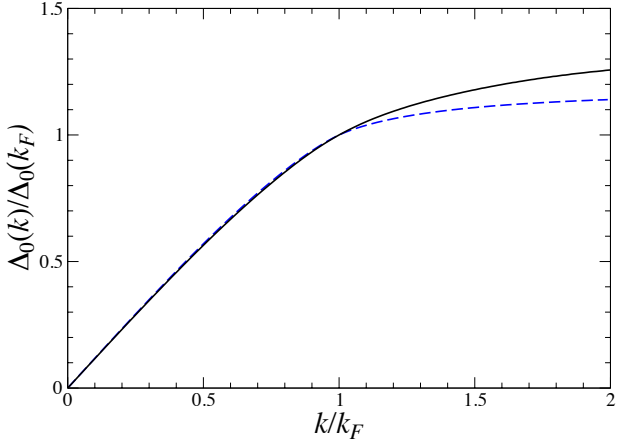}
\caption{(color online).
The zero-temperature gap $\Delta_0(k)$ in units of $\Delta_0(k_F)$ as a function of $k/k_F$ for $\Omega=0.25$, $\beta=0$, and $k_Fr_*=0.4$ near the minimum of inelastic
losses ($r_\delta/\lambda_\delta=10.5$).  The solid curve is the result of full scale numerics on the basis of Eq.~(\ref{gappwave1}), and the dashed curve is the result of
Eq.~(\ref{DkF}).
\label{fig:delta}}
\end{figure}

Equation (\ref{DeltakkF}) can be used to calculate the contribution to the integral in Eq.~(\ref{gappwave1}) from momenta $k'$ away from the Fermi surface. This immediately
leads to a relation between the zero-temperature order parameter on the Fermi surface, $\Delta_0(k_F)$, and the critical temperature $T_c$. Putting $k=k_F$ in Eq.~(\ref{gappwave1})
and taking into account that in the weak coupling regime one has $|\Delta(k)|,T_c\ll E_F$, we divide the region of integration over $k'$ into two parts: $|E_{k'}-E_F|<\omega$, and
$|E_{k'}-E_F|>\omega$, where $|\Delta(k)|,T_c\ll\omega\ll E_F$. In the second region we may put $\epsilon_{k'}=|\xi_{ k'}|$ and $\tanh(\epsilon_{k'}/2T)=1$, so that ${\cal K}
(k')=1/2|\xi_{k'}|$. Then, making use of Eq.~(\ref{DeltakkF}) for $\Delta(k')$ we see that the integration in this region gives $\Delta(k_F){\cal A}$, where the quantity ${\cal A}$
is temperature independent. Dividing both sides of the gap equation by $\Delta(k_F)$ we obtain:
\begin{eqnarray}       
\!\!\!&&-\frac{m}{2\pi\hbar^2}\!\int_0^{\omega}\!\!\![{\bar f}(k_F)\!+\!\delta V(k_F,k_F)]\!\tanh\!\left(\frac{\sqrt{\xi_{k'}^2\!+\!|\Delta(k_F)|^2}}{2T}\!\right)  \nonumber \\
\!\!\!&&\times\frac{d\xi_{k'}}{\sqrt{\xi_{k'}^2+|\Delta(k_F)|^2}}\,+{\cal A}=1. \label{calA}
\end{eqnarray}
As $T\rightarrow T_c$ the order parameter tends to zero and we may put $\Delta(k_F)=0$ in Eq.~(\ref{calA}). Then, subtracting this equation at $T\rightarrow T_c$ from
the equation at $T=0$ we have:
\begin{equation*}
\int_0^{\omega}\left\{\frac{\tanh(\xi_{k'}/2T_c)}{\xi_{k'}}-\frac{1}{\sqrt{\xi_{k'}^2+|\Delta_0(k_F)|^2}}\right\}d\xi_{k'}=0.
\end{equation*}
The integral converges at $\xi_{k'}$ of the order of $T_c$ or $|\Delta_0(k_F)|$, and we may extend the upper limit of integration to infinity. This gives:
\begin{equation}         \label{standard}
T_c=\left(\frac{e^{C}}{\pi}\right)|\Delta_0(k_F)|,
\end{equation}
which is the same relation as in the case of the $s$-wave pairing in a two-component weakly interacting 3D Fermi gas \cite{LL9}. Note that equation (\ref{calA}) allows one to 
establish a relation between $T_c$ or $|\Delta_0(k_F)|$ and $|\Delta(k_F)|$ at any temperature. Having in mind that the momentum dependence of the order parameter follows from
Eq.~(\ref{DeltakkF}) we see that the calculation of $T_c$ provides us with the order parameter at any $k$ and temperature.    

We now calculate the critical temperature $T_c$ on the basis of the gap equation (\ref{gappwave1}) at $k=k_F$. In the first line of this equation we again divide the region of
integration into two parts: $|E_{k'}-E_F|<\omega$, and $|E_{k'}-E_F|>\omega$. In the first region we put $\Delta(k')=\Delta(k_F)$ and ${\bar f}(k',k_F)={\bar f}(k_F)$. The
contribution of the second term in square brackets is then equal to zero. In the first term we set $\epsilon_{k'}=|\xi_{k'}|$ and, hence, ${\cal K}(k')=(2|\xi_{k'}|)^{-1}\tanh(
|\xi_{k'}|/2T_c)$. Denoting the result of the integration of this term as $\Delta_1(k_F)$ we have:
\begin{eqnarray}           
&&\Delta_1(k_F)=\Delta(k_F)\frac{4k_Fr_*}{3\pi}\left(1-\frac{3\pi}{16}k_Fr_*\ln(\rho k_Fr_*)\right)   \nonumber \\
&&\times\ln\left(\frac{2e^C\omega}{\pi T_c}\!\right), \label{Delta1}
\end{eqnarray}
where we used ${\bar f}(k_F)={\bar f}_1(k_F)+{\bar f}_2(k_F)$, with
${\bar f}_1,\,{\bar f}_2$ given by equations (\ref{f1k}), (\ref{f2k})
and the numerical factor $\rho$ introduced after
Eq.~(\ref{deltafinal}).

In the second region, $|E_{k'}-E_F|>\omega$, we put ${\cal K}(k')=1/2|\xi_{k'}|$. The contribution of this region is small as $\sim k_Fr_*$ compared to the result of
Eq.~(\ref{Delta1}). Therefore, it is sufficient to retain only the leading low-momentum contribution to the off-shell amplitude ${\bar f}(k',k_F)$. Thus, using Eq.~(\ref{f1kk})
for ${\bar f}(k',k_F)$ and Eq.~(\ref{DkF}) for $\Delta(k')$ we write this contribution as
\begin{eqnarray}       
\Delta_2(k_F)=\Delta(k_F)\frac{3\pi r_*}{8k_F}\int_0^{k_{\omega}}\!\!\!\frac{k'^3dk'}{(k_F^2-k'^2)}F^2\left(-\frac{1}{2},\frac{1}{2},2,\frac{k'^2}{k_F^2}\right), \nonumber
\end{eqnarray}
where $k_{\omega}=\sqrt{2m(E_F-\omega)/\hbar^2}$. Taking into account that $\omega\ll E_F$ we then obtain:
\begin{equation}          \label{Delta2}
\Delta_2(k_F)=\Delta(k_F)\frac{4k_Fr_*}{3\pi}\left[\ln\left(\frac{E_F}{\omega}\right)-\eta\right],
\end{equation}
where
\begin{eqnarray}
\eta=&&\!\!\!\!1-\frac{9\pi^2}{64}\int_0^1\left[F^2\left(-\frac{1}{2},\frac{1}{2},2,x\right)-F^2\left(-\frac{1}{2},\frac{1}{2},2,1\right)\right]  \nonumber \\
&&\times\frac{xdx}{1-x}\simeq 0.78. \nonumber
\end{eqnarray}

The main contribution to the term in the second line of Eq.~(\ref{gappwave1}), describing Gor'kov-Melik-Barkhudarov corrections, comes from the vicinity of the Fermi surface, i.e.
from the region where $|E_{k'}-E_F|<\omega$. The result of the integration reads:
\begin{equation}         \label{Delta3}
\Delta_3(k_F)=-\Delta(k_F)\frac{\delta V(k_F,k_F)}{2\pi}\ln\left(\frac{2e^C\omega}{\pi T_c}\right).
\end{equation}
In the Appendix we show that in the low-momentum limit one has
\begin{equation}     \label{deltaVF}    
\delta V(k_F,k_F)=\alpha\frac{\hbar^2}{m} (k_Fr_*)^2, 
\end{equation}
where $\alpha\simeq 2.3$. 

Then, making a summation of the contributions (\ref{Delta1}),
(\ref{Delta2}), and (\ref{Delta3}), and dividing the gap equation by
$\Delta(k_F)$ we obtain:
\begin{eqnarray}
&&\frac{4k_Fr_*}{3\pi}\left[\ln\left(\frac{2e^CE_F}{\pi T_c}\right)-\eta\right]- (k_Fr_*)^2\ln\left(\frac{E_F}{T_c}\right) \nonumber  \\
&&\times\left\{\frac{1}{4}\ln(\rho k_Fr_*)+\frac{\alpha}{2\pi}\right\}=1.  \label{findTc}
\end{eqnarray}
Note that we put $\omega\sim E_F$ in the argument of the logarithm in the Gor'kov-Melik-Barkhudarov term and in the term proportional to $(k_Fr_*)^2\ln(\rho k_Fr_*)$. This is
justified because $\omega$ can be chosen as a small numerical fraction of $E_F$, and these terms are small as $\sim k_Fr_*$ compared to the leading term. 

We now recall that the bare mass $m$ should be replaced by the effective mass $m_*$ following from Eq.~(\ref{meff}). The relative difference between $m_*$ and $m$ is proportional
to
$k_Fr_*$ and is small. Therefore it is sufficient to replace $m$ with $m_*$ only in the multiple $r_*\propto m$ in the first term of Eq.~(\ref{findTc}). Using Eq.~(\ref{meff}) this
leads to the appearance of a new term $(16/9\pi^2)(k_Fr_*)^2\ln(E_F/T_c)$ in equation (\ref{findTc}), which is equivalent to replacing $\alpha$ by
$\tilde\alpha=\alpha-32/9\pi\simeq 1.17$. Relying on the
inequality $k_Fr_*\ll 1$ we then immediately find the transition
temperature \cite{tcnote}:
\begin{equation}            \label{Tcfin}
\frac{T_c}{E_F}=\frac{\kappa}{(k_Fr_*)^{9\pi^2/64}}\exp\left(-\frac{3\pi}{4k_Fr_*}\right),
\end{equation}
where
$$\kappa=\exp\left\{-\frac{9\pi^2}{64}\ln\rho-\frac{9\pi}{32}\tilde\alpha+C-\eta+\ln\left(\frac{2}{\pi}\right)\right\}.$$
Substituting the values of $\rho$, $\eta$, and $\tilde\alpha$ specified above we obtain:
\begin{equation}         \label{kappa}
\kappa\simeq 0.16\exp\left(-\frac{9\pi^3 A}{64}\right).
\end{equation}

Let us make several important statements regarding 
equation (\ref{Tcfin}). First of all, the factor in the exponent is
$\sim -(k_Fr_*)^{-1}$, and the ratio $T_c/E_F$ can be made larger than
$10^{-2}$ even for relatively small $k_Fr_*$. This is a consequence of
the anomalous scattering due to the attractive $1/r^3$ tail of the
potential $V_0 (r)$, which gives a scattering amplitude proportional
to $k$. Moreover, an accurate calculation of the scattering amplitude
reveals a $(kr_*)^2\ln (kr_*)$ term, which leads to the appearance
of a power law factor $(k_Fr_*)^{-9\pi^2/64}$ in front of the exponent
and provides an additional increase of the critical temperature. This
behavior is in contrast to the $p$-wave pairing of short-range
interacting atoms, where the factor in the exponent is inversely
proportional to $-k_F^2$ and the BCS critical temperature is
vanishingly low.

The dependence of $T_c$ on the short-range behavior of the potential
$V_0(r)$ is contained in the factor $\kappa$ through the coefficient
$A$. In Fig.~8 we present $A$ and $\kappa$ as functions of
$r_{\delta}/\lambda_{\delta}$ for $\Omega_R=0.25\delta$ and $\beta$
ranging from $0$ to $0.2$. These results show that $\kappa$ can be
varied by two orders of magnitude by changing
$r_{\delta}/\lambda_{\delta}$ [{\em i.e.} by changing the depth of the
potential $V_0(r)$] while remaining in the regime without bound states
in this potential, such that a rapid three-body decay is absent.

\begin{figure}[ttp]
\includegraphics[width=1.0\columnwidth]{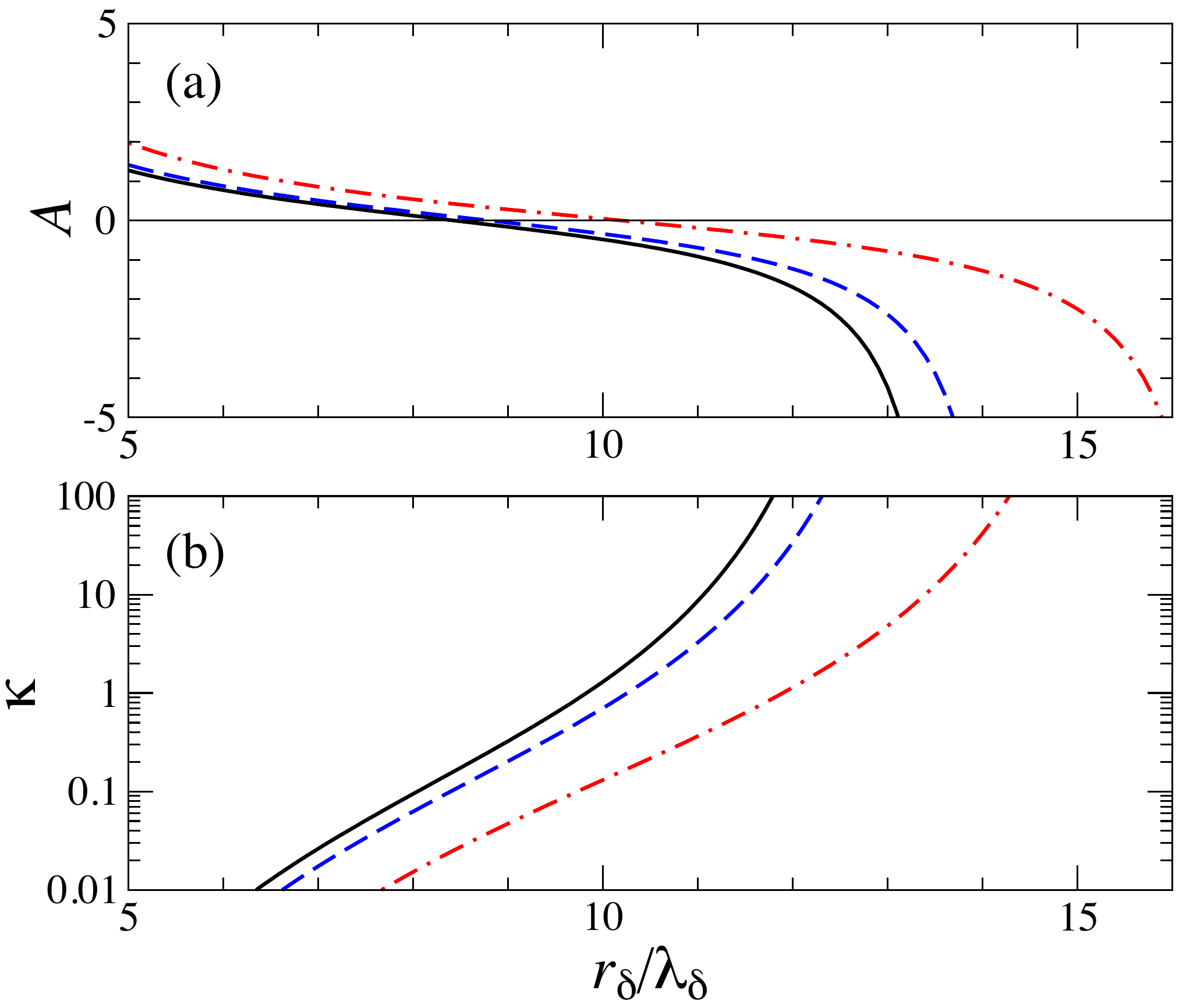}
\caption{(color online). The coefficients $A$ and $\kappa$ versus
  $r_\delta/\lambda_\delta$ for $\Omega_R/\delta=0.25$ and $\beta=0$
  (black, solid), 0.1 (blue, dashed), and 0.2 (red, dash-dotted). The
  coefficients are presented for the values of
  $r_\delta/\lambda_\delta$ at which bound states in the potential
  $V_0(r)$ do not exist.}
\end{figure}

In Fig.~9 we display the ratio $T_c/E_F$ as a function of $k_Fr_*$ for
$\Omega_R=0.25\delta$ and $r_{\delta}/\lambda_{\delta}$ close to the
minimum of the loss rate at a given $\beta$. We compare the full scale
numerical solution of the gap equation (\ref{gappwave1}) with the
analytic expression, Eq.~(\ref{findTc}).  The discrepancy between
analytic and numerical results as $k_Fr_*$ approaches 1 is to be
expected as the analytic method is perturbative in this
parameter. From the results of Fig.~9 we conclude that a critical
temperature of the order of $5\%$ of the Fermi energy is realistic.
For typical 2D densities $n\sim 10^8 - 10^9$ cm$^{-2}$ the Fermi
energy of alkali atom molecules is of the order of hundreds of
nanokelvins. Then, as we see from Fig.~9, the superfluid transition
temperature can be as high as $10$ or $20$ nK.


\begin{figure}[ttp]
\includegraphics[width=1.0\columnwidth]{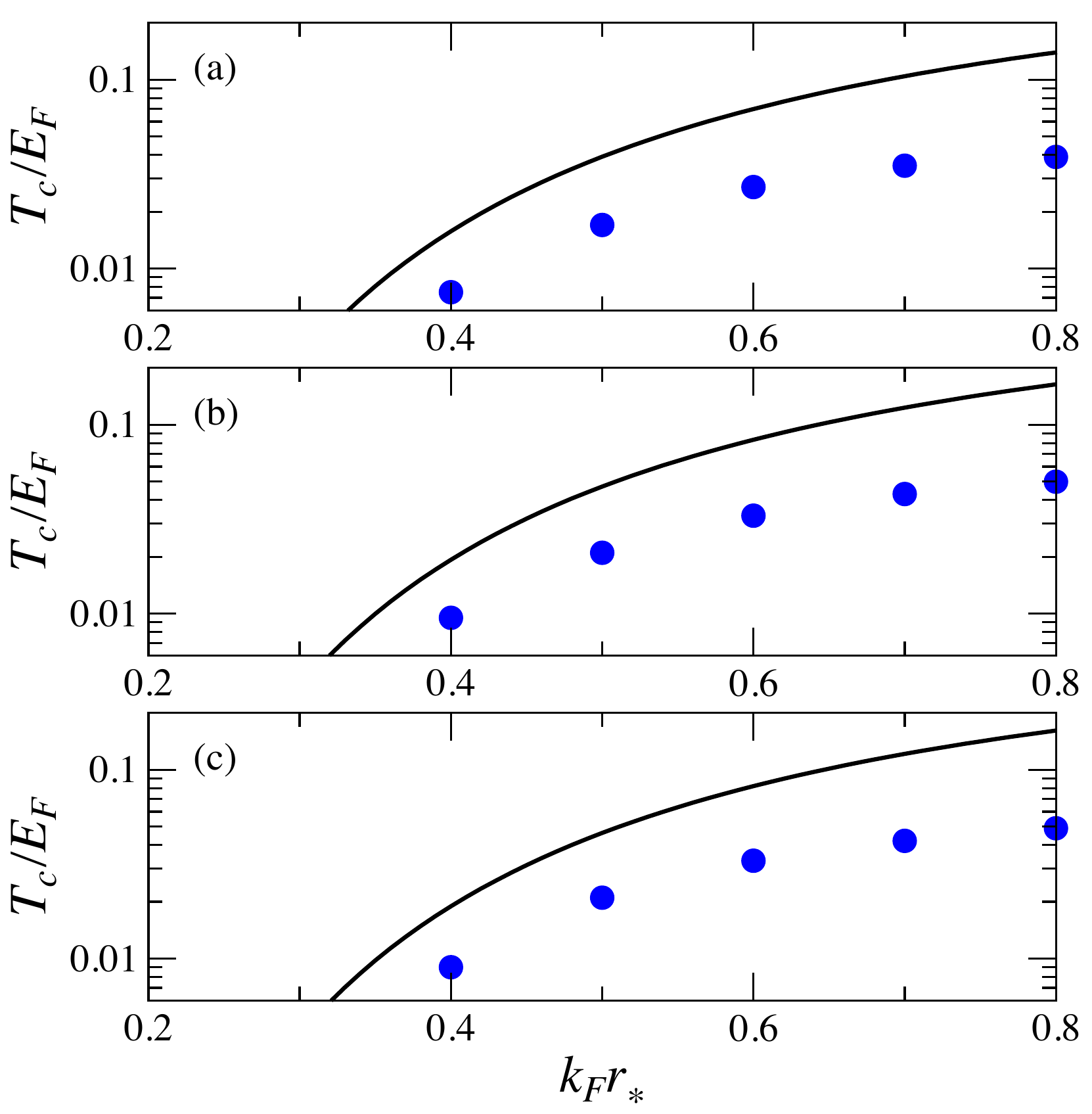}
\caption{(color online). Critical temperature $T_c$ in units of the
  Fermi energy as a function of $k_Fr_*$ for $\Omega_R/\delta=0.25$ near the
  minimum of inelastic losses. In (a) $\beta=0$,
  $r_\delta/\lambda_\delta=10.7$, in (b) $\beta=0.1$,
  $r_\delta/\lambda_\delta=11.4$, and in (c) $\beta=0.2$,
  $r_\delta/\lambda_\delta=13.2$. The data points are obtained from
  full scale numerics on the basis of Eq.~(\ref{gappwave1}), and the
  curves from Eq.~(\ref{findTc}) \cite{tcnote}.
  \label{fig:tcoveref}}
\end{figure}

\section{Concluding remarks}

The results described in this paper indicate that it is realistic to
create the superfluid topological $p_x+ip_y$ phase with alkali atom
polar molecules. The key point is the anomalous scattering by the
attractive $1/r^3$ tail of the interaction potential $V_0(r)$, which
leads to a relatively large (and negative) $p$-wave scattering
amplitude and thus to an achievable transition temperature $T_c$ even
in the BCS regime. What we have shown in the present paper is the
possibility to manipulate $T_c$ by modifying the short-range part of
the potential $V_0(r)$. As is seen from Fig.~8, the pre-exponential
coefficient $\kappa$ can be varied within 2 or 3 orders of magnitude,
and we may still remain in the BCS regime with fairly small inelastic
losses.

Another result of this paper is that the addition of a {\em dc} electric field reduces the inelastic losses, and one can work at higher densities. Consider, for example, $^6$LiCs
molecules which have a permanent dipole moment of $5.5$ D \cite{Dulieu,Weidemuller}. For $\Omega_R=0.12\delta$ and $\beta=0.165$ the dipole-dipole distance is $r_*=100$ nm and
$k_Fr_*=1$ at 2D densities of $10^9$ cm$^{-2}$ corresponding to a Fermi energy of 200 nK. Then,  our calculations give the rate constant $~10^{-9}$ cm$^2$/s for
$r_{\delta}/\lambda_{\delta}=39$, which for the selected $\Omega_R/\delta$ and $\beta$ corresponds to the absence of bound states in the potential $V_0(r)$ and is not far from the
minimum of inelastic losses. Thus, the lifetime of the gas is about 1 s at these densities.   
With $k_Fr_*=1$ we are beyond the BCS regime. However, the results of Fig.~9 indicate that in this case we are likely to have a critical temperature of at least a twentieth part
of $E_F$, which is $T_c\simeq 10$ nK. Note that the ratio $r_{\delta}/\lambda_{\delta}=39$ corresponds to $r_{\delta}\simeq 15$ nm and using the rotational constant $B\simeq 0.27$
K \cite{Weidemuller} we find that $\beta=0.165$ corresponds to $E_{dc}\simeq 350$ V/cm. 

In order to provide the 2D regime of scattering and superfluid pairing one should have the confinement length in the $z$-direction, $l_z\lesssim r_{\delta}$. For the above example
of $^{6}$LiCs molecules with $r_{\delta}\simeq 15$ nm this requires a confinement frequency $\omega_z=\hbar/ml_z^2$ of the order of 300 kHz. Such very high frequencies are not
unrealistic and are achievable by a tight optical confinement. In particular, frequencies close to 100 kHz have been used for sideband cooling of cesium atoms
\cite{Vuletic,Bouchoule,Bouchoule2}. 

Similar estimates are obtained for LiK and NaK molecules which have dipole moments of $3.5$ and $2.7$ D, respectively \cite{Dulieu}. Here one can have the Fermi energy above 300 nK
($n\gtrsim 10^9$ cm$^{-2}$) and $T_c$ on the level of tens of nanokelvins, with a lifetime of the order of a second. However, the necessary confinement frequencies are about 500
kHz.

The requirement for the confinement frequency can be relaxed by introducing a shallow optical lattice and thus increasing the effective mass
of the molecules. For example, increasing the effective mass by a factor of 5 allows one to increase $r_{\delta}$ to 25 nm for $^6$LiCs molecules, which corresponds to the
confinement frequency of 100 kHz. Then, however, we have to have larger $r_*$ and even for $k_Fr_*\sim 1$ the Fermi energy is about 20 nK, so that $T_c$ will be on the level of
nanokelvins. Similar estimates are obtained for LiK and NaK molecules where under the same increase of the effective mass the confinement frequency can be decreased to 200 kHz,
with $T_c$ of the order of nanokelvins. Note that increasing the effective mass by a factor of 10 makes the situation promising even for $^{40}$KRb molecules which have a much
smaller dipole moment
$d=0.6$ D \cite{Ni2008}. Then, taking $r_{\delta}\simeq 20$ nm, which
corresponds to the confinement frequency of 200 kHz and to $r_*=100$ nm we find $T_c$ on the level of
nanokelvins ($E_F\simeq 30$ nK) at densities of $10^9$ cm$^{-2}$ and with a lifetime of the order of seconds. 

Note that the strong confinement in the $z$-direction with frequencies
$\omega_z$ exceeding 100 kHz, originating from the condition
$l_z\lesssim r_{\delta}$, is only necessary for considering the
relative motion of colliding molecules as two-dimensional at
intermolecular distances $r\sim r_\delta$ and revealing the role of
short-range physics for the elastic $p$-wave interaction and inelastic
processes. The 2D superfluid pairing by itself, which leads to the
$p_x+ip_y$ symmetry of the ground state and mostly originates from the
anomalous scattering, simply requires the dipole-dipole distance $r_*$
significantly larger than $l_z$. For all examples given above, this
condition is well satisfied at a frequency $\omega_z\approx 50$
kHz. However, the absence of the inequality $r_\delta\lesssim l_z$
introduces a numerical uncertainty in the coefficient $\kappa$ in
Eq. (61) and can make the inelastic decay faster.

The formation of the $p_x+ip_y$ superfluid phase should be apparent in
numerous observables.  Superfluidity itself can be detected by the
means that have been used to detect $s$-wave
superfluids \cite{Varenna,Trento}. The most interesting new properties arise
in the presence of quantized vortices, which are predicted to carry
zero energy Majorana modes on their cores \cite{Green}. Vortices can be
induced in superfluid Fermi gases by causing the gas to
rotate \cite{Zwierlein,Varenna}.  The appearance of zero energy
Majorana modes on the vortex cores will lead to signatures in the RF
absorption spectrum: the Majorana modes are predicted to lead to a
series of sharp lines below the superfluid gap. These lines arise from
processes in which a fermion is excited out of the Majorana modes
\cite{grosfeld:104516}. Ultimately one would hope to probe non-abelian
exchange statistics of these vortices \cite{tewari-2007-98}.

\section*{Acknowledgements}

We acknowledge support from EPSRC Grant No. EP/F032773/1, from the IFRAF Institute, and from the Dutch Foundation FOM. One of us (J.L.) acknowledges support by a Marie Curie
  Intra European Fellowship within the 7th European Community
  Framework Programme. This research has been supported in part by the National
Science Foundation under Grant No. NSF PHYS05-51164. LPTMS is a mixed research unit No. 8626 of CNRS and Universit\'e Paris Sud.



\section*{Appendix}

In this Appendix, we demonstrate how the Gor'kov-Melik-Barkhudarov
corrections to the bare interparticle interaction may be computed. The
corrections consist of the four diagrams depicted in Fig.~6
\cite{Baranov2002}. In the low-momentum limit they are dominated by
the long-range $1/r^3$ tail of the potential $V_0(r)$ and, hence,
we may approximate this potential as
\begin{equation}
  V_0(r)\approx
\left\{\begin{array}{ll}0, \hspace{7mm} & r< r_0\\
-(\hbar^2/m)r_*/r^3, & r>r_0
\end{array},
\right.
\end{equation}
where $r_0$ is much smaller than the de Broglie wavelength of
particles. The Fourier transform of the potential takes the form:
\begin{equation}
V_0({\bf q})\approx -\frac{2\pi\hbar^2}m\frac{r_*}{r_0}+
\frac{2\pi\hbar^2}m |{\bf q}|r_*.
\label{eq:Vft}
\end{equation}
The expressions for the four diagrams of Fig. 6 are
\begin{eqnarray}
\delta V_{a}({\bf p}',{\bf p}) &=&\int \frac{d^2q}{(2\pi )^{2}}
\frac{N({\bf q}+{\bf p}_{-}/2)-N({\bf q}-{\bf p}_{-}/2)}{\xi _{{\bf q}+{\bf p
}_{-}/2}-\xi _{{\bf q}-{\bf p}_{-}/2}}\nonumber\\ &&\times
V_{0}^{2}({\bf p}_{-}), \label{eq:fig6a}
\end{eqnarray}
\begin{eqnarray}
\delta V_{b}({\bf p}',{\bf p}) &=&-\int \frac{d^2q}{(2\pi )^{2}}
\frac{N({\bf q}+{\bf p}_{-}/2)-N({\bf q}-{\bf p}_{-}/2)}{\xi _{{\bf q}+{\bf p
}_{-}/2}-\xi _{{\bf q}-{\bf p}_{-}/2}}\nonumber\\ &&\times V_{0}({\bf p}_{-})V_{0}({\bf q}-{\bf p}
_{+}/2), 
\end{eqnarray}
\begin{eqnarray}
\delta V_{c}({\bf p'},{\bf p}) &=&-\int \frac{d^2q}{(2\pi )^{2}}
\frac{N({\bf q}+{\bf p}_{-}/2)-N({\bf q}-{\bf p}_{-}/2)}{\xi _{{\bf q}+{\bf p
}_{-}/2}-\xi _{{\bf q}-{\bf p}_{-}/2}}\nonumber\\ &&\times V_{0}({\bf p}_{-})V_{0}({\bf q}+{\bf p}
_{+}/2),
\end{eqnarray}
\begin{eqnarray}
\delta V_{d}({\bf p}',{\bf p}) &=&-\int \frac{d^2q}{(2\pi )^{2}}
\frac{N({\bf q}+{\bf p}_{+}/2)-N({\bf q}-{\bf p}_{+}/2)}{\xi _{{\bf q}+{\bf p
}_{+}/2}-\xi _{{\bf q}-{\bf p}_{+}/2}}\nonumber\\ &&\times V_{0}({\bf q}-{\bf p}_{-}/2)V_{0}({\bf 
q}+{\bf p}_{-}/2), \label{eq:fig6d}
\end{eqnarray}
with ${\bf p}_{\pm}={\bf p}'\pm{\bf p}$, and $\xi_{\bf q}=\hbar^2{\bf
  q}^2/2m-\mu$. In the weakly interacting regime we may put the chemical potential $\mu$ equal to the Fermi
energy, and at very low temperatures $T<T_c$ we may take the step function $N({\bf q})=\Theta(k_F-|{\bf q}|)$ for the Fermi-Dirac distribution.

The $p$-wave part of the diagrams (\ref{eq:fig6a})-(\ref{eq:fig6d}) is
\begin{eqnarray}
\!\!\delta V(p',p)\!\!=\!\!\sum_j\delta V_j(p',p)\!\!=\!\!\!\int_0^{2\pi}\!\frac{d\phi}{2\pi}
e^{-i\phi}\!\!\!\!\!\!\!\sum_{j\in\left\{a,b,c,d\right\}}\!\!\!\!\!\!\!\!\delta V_j({\bf p}',{\bf p}),
\end{eqnarray}
where $\phi$ is the angle between ${\bf p}'$ and ${\bf p}$.  One can easily check that the momentum-independent term of $V_0({\bf q})$ does not contribute to the sum of the four
diagrams.

For the analytical calculation of $T_c$ we only need the contributions
$\delta V_j(p',p)$ on the Fermi surface, i.e. for $p'=p=k_F$. In this
case each contribution can be represented in the form $\delta
V_j(k_F,k_F)=\alpha_j(\hbar^2/m)(k_Fr_*)^2$. The calculation of the
diagram \ref{fig:gm}(a) is straightforward and it gives
$\alpha_a=2\pi$. The other diagrams we calculated numerically. The
values of the coefficients are
\begin{equation}
\alpha_a=2\pi,\hspace{5mm}
\alpha_b=\alpha_c\simeq-1.5,\hspace{5mm}
\alpha_d\simeq-1.0,
\end{equation}
and thus we find the result of Eq.~(\ref{deltaVF}):
\begin{equation}
\delta V(k_F,k_F) \equiv \alpha
\frac{\hbar^2}m\left(k_Fr_*\right)^2,
\end{equation}
with $\alpha\simeq2.3$.

For the numerical solution of the gap equation we used a simplified
dependence of $\delta V$ on $p'$ and $p$, namely the dependence
following from the diagram (a) with $V_0(|{\bf
  p}_-|)=(2\pi\hbar^2/m)|{\bf p}_-|r_*$. This dependence is given by:
\begin{eqnarray}
&&\hspace{-10mm}\delta V(p',p)=\delta V(k_F,k_F)
\Big\{\frac{pp'}{k_F^2}+  \nonumber \\
&&\int_0^{2\pi}\frac{d\phi}{2\pi}\Theta(p_--2k_F)\cos\phi\frac{p_-\sqrt{p_-^2-4k_F^2}}{k_F^2}\Big\}.
\end{eqnarray}
For $(p+p')<2k_F$ we have $\delta V(p',p)=\delta V(k_F,k_F)pp'/k_F^2$, and this quantity rapidly decays at larger momenta.

\end{document}